\documentclass[12pt]{article}

\usepackage[margin=1.25in]{geometry}
\usepackage{setspace}
\usepackage{enumitem}
\usepackage{mathtools}
\usepackage{dsfont}
\usepackage{amsthm}
\usepackage{amssymb}
\usepackage{bm}
\usepackage{amsfonts} 
\usepackage{graphicx}
\usepackage{float}
\usepackage[bottom]{footmisc}
\usepackage[titletoc]{appendix}
\usepackage{caption}
\usepackage[flushleft]{threeparttable}
\allowdisplaybreaks
\usepackage{csquotes}
\usepackage{wrapfig}
\usepackage{pgfplots,pgfplotstable}
\pgfplotsset{compat=1.11} 
\usepackage{tikz}
\usetikzlibrary{patterns,decorations.pathreplacing}
\usepackage[authoryear]{natbib}
\usepackage[hidelinks]{hyperref}
\everymath{\displaystyle}

\newtheorem{assumption}{Assumption}
\newtheorem{lemma}{Lemma}
\newtheorem{theorem}{Theorem}

\newtheoremstyle{named}{}{}{\itshape}{}{\bfseries}{.}{.5em}{#1\thmnote{ #3}}
\theoremstyle{named}
\newtheorem*{namedassumption}{Assumption}

\newtheoremstyle{remark}{}{}{}{}{\bfseries}{.}{.5em}{#1\thmnote{ #3}}
\theoremstyle{remark}
\newtheorem*{namedexample}{Example}

\newcommand{\inv}{^{-1}}

\newcommand{\argmin}{\textrm{ \emph{argmin} }}

\newcommand{\R}{\mathbb{R}}

\newcommand{\shrinkageparameter}{1}

\numberwithin{equation}{section}

\begin{document}

\title{
Weak Identification with Bounds\\ in a Class of Minimum Distance Models
}
\author{
Gregory Fletcher Cox\footnote{Department of Economics, National University of Singapore, 
\href{mailto:ecsgfc@nus.edu.sg}{ecsgfc@nus.edu.sg}}\\
\today
}
\date{}
\maketitle
\vspace{-0.5cm}

\begin{abstract}
When parameters are weakly identified, bounds on the parameters may provide a valuable source of information. 
Existing weak identification estimation and inference results are unable to combine weak identification with bounds. 
Within a class of minimum distance models, this paper proposes identification-robust inference that incorporates information from bounds when parameters are weakly identified. 
This paper demonstrates the value of the bounds and identification-robust inference in a simple latent factor model and a simple GARCH model. 
This paper also demonstrates the identification-robust inference in an empirical application, a factor model for parental investments in children. 
\end{abstract}



\section{Introduction}

Minimum distance models are commonly used to estimate and test hypotheses on structural parameters of interest. 
The link function, which maps the structural parameters to reduced-form parameters, is usually nonlinear and may not be one-to-one. 
If the link function is not one-to-one, the structural parameters are not identified. 
Without identification, the minimum distance estimator is inconsistent, and standard hypothesis tests are invalid. 
Furthermore, weak identification arises in a neighborhood of parameter values that are not identified. 

When identification fails, information about the structural parameters may still be available from bounds on the identified set. 
This paper combines weak identification with bounds by (1) formulating limit theory that accommodates an identified set that is defined by an intersection of bounds, (2) proposing a weak-identification-robust (WIR) test that has power against hypotheses that violate the bounds when identification is weak, and (3) demonstrating this test in latent factor models and an empirical application. 
Existing WIR tests are either invalid in the presence of bounds or do not have power against hypotheses that violate the bounds. 

\textbf{Contribution (1) Limit Theory.} 
We formulate limit theory that accommodates an identified set that is defined by an intersection of bounds. 
The main challenge is relaxing a product-space assumption on the parameter space that is common in the weak identification literature. 
We allow the parameter space to take a general shape. 
This requires careful modifications to existing arguments. 
In particular, this paper uses a novel argument involving a shrinking sequence of neighborhoods of the identified set combined with an appeal to a new argmax theorem developed in \cite{CoxArgmax}. 

The weak identification literature classifies different types of limit theory: strong, semi-strong, and weak, indicating the strength of identification. 
The presence of the bounds influences each type of limit theory. 
Under weak identification, the limit is characterized by the argmin of a stochastic process over the identified set, which is circumscribed by the bounds. 
Under strong identification, the limit is the argmin of a stochastic process over a setwise limit of the local parameter space, allowing the parameter to be on or near the boundary. 
Under semi-strong identification, the limit is the same as under strong identification, except the influence of the bounds, through the setwise limit of the local parameter space, is rotated so the bounds only constrain the semi-strongly identified parameters. 
The semi-strong identification limit also has the feature that the true parameter may be further from the boundary (the sequence of true parameter values may converge to the boundary arbitrarily slowly) and yet the boundary still influences the limit. 
This is a novel feature of the limit theory that is unique to the combination of weak identification with bounds. 
It must be carefully addressed to achieve uniformly valid inference. 

\textbf{Contribution (2) WIR Inference.} 
The limit theory is used to define a WIR test. 
When testing a hypothesis on the structural parameters, we recommend calculating the quasi-likelihood ratio (QLR) statistic and comparing it to a critical value that is calculated from the quantiles of the limiting distribution under weak and strong identification. 
Interestingly, the quantiles of the limiting distribution under semi-strong identification are not needed. 
We show that, if the parameters indexing the limit theory are estimated in the right way, then the estimated quantiles of the weak and strong limits adapt to the semi-strong limit. 

This robust QLR (RQLR) test is very flexible, applying to full-vector inference and subvector inference and allowing strongly identified or weakly identified nuisance parameters. 
It has two noteworthy features. 
First, the test is asymptotically valid uniformly over the strength of identification and the proximity of the boundary. 
Second, simulations show the test has power against hypotheses that violate the bounds when identification is weak. 
In contrast, other WIR tests do not share these noteworthy features. 
Surprisingly, this includes tests based on the Anderson-Rubin (AR) statistic from \cite{StockWright2000} and the K and CLR statistics from \cite{Kleibergen2005}.\footnote{In \cite{StockWright2000}, the AR statistic is denoted ``S'' and in \cite{Kleibergen2005}, the CLR statistic is denoted ``GMM-M.''} 
Specifically, the versions of these tests that plug in estimates of \textit{strongly identified} nuisance parameters either do not have power against hypotheses that violate the bounds or do not control size for hypotheses near the boundary of the identified set.\footnote{Controlling size near the boundary of the identified set is especially important because, otherwise, a confidence set formed by inverting such a test is too small and under-covers asymptotically. Likewise, having consistent power against hypotheses that violate the bounds is also important because then a confidence set formed by inverting such a test is smaller.} 
It is well-known that, in general, one cannot plug in \textit{weakly identified} nuisance parameters for these tests. 
The fact that plugging in \textit{strongly identified} nuisance parameters is problematic near the boundary of the identified set is novel and is documented in the simulations in Section 5. 

\textbf{Contribution (3) Examples and Application.} 
There are many possible applications of weak identification with bounds in minimum distance models. 
We demonstrate the theory in two simple examples: a latent factor model with one factor and a GARCH model. 
In the factor model, weak identification arises when the factor loadings are close to zero. 
However, informative bounds on the weakly identified parameters can be derived from the nonnegativity of the variances of the latent factor and errors. 
In the GARCH model, weak identification arises when the ARCH coefficient is close to zero. 
We show how to reparameterize both models to fit the required weak identification structure and detail implementation instructions for the RQLR test. 

Weak identification may be relevant in many empirical applications of factor models, especially when the number of factors is unknown. 
If a researcher does not know whether a given dataset contains one or two factors, for example, then the possibility the dataset contains a weakly identified factor should not be ruled out. 
We consider one such application, to estimating a model of parental investments in children. 
\cite{Attanasio2020AER} use a factor model with one factor to analyze the effects of a randomized intervention designed to improve the quality of maternal-child interactions on parental investments in children. 
Following \cite{CoxWeakIdFactor}, we note indicators that the model contains a second factor that may be weakly identified. 
We demonstrate the relevance of the bounds and the value of inference based on the RQLR test in a weakly identified factor model with two factors. 

\textbf{Related Literature.} 
Broadly speaking, there are two literatures in econometrics that are designed to cover models without identification: set-identified models (including moment inequality models) and weakly identified models (including weak instruments). 
Generally, the models covered by these literatures are distinct. 
Set-identified models are designed to cover the case that the identified set is (setwise) continuous as a function of the true value of the parameter. 
In contrast, weakly identified models are designed to cover a particular type of discontinuity in the identified set---one where the identified set is usually a singleton but becomes a non-singleton set on a subspace of the parameter space. 

To clarify, there is an important exception where the set-identified literature covers weak identification, which is hypothesis testing on the full vector of parameters, for example in \cite{AndrewsSoares2010}. 
In that case, one can use test statistics that plug in the value of the parameters under the null and are therefore insensitive to the Jacobian of the moments being close to singular. 
For this reason, the problem of weak identification manifests in papers on inference for subvectors or functions of the parameters in the set-identified literature. 
We briefly discuss the main papers in that literature with a focus on their capacity to cover weak identification. 
Assumption A.3(a) in \cite{BugniCanayShi2017} and Assumption 2 in \cite{Bei2024} require, as a necessary condition, the identified set to be continuous as a function of the true distribution of observables, which rules out weak identification. 
\cite{AndrewsRothPakes2023}, \cite{CoxShi2023}, and \cite{FSST2023} focus on moments that are linear in the parameters, and, even then, require the Jacobian to be known so it cannot be close to singular.\footnote{\cite{CoxShi2023} allows the Jacobian to be close to singular, but their test is not valid for parameter values in the identified set at the limit of the sequence; it is only valid for sequences of parameter values in the sequence of identified sets. This makes a difference when the identified set is (setwise) discontinuous.} 
\cite{KaidoMolinariStoye2019} can cover weak identification, at least in theory, because of a rho-box construction designed to avoid assumptions like Assumption A.3(a) in \cite{BugniCanayShi2017}. 
\cite{KaidoMolinariStoye2019} can be viewed as an attempt to bridge the gap between the weak identification and set-identification literatures. 
This paper also attempts to bridge that gap but starting from the weak identification side. 
We include the recommended inference methods from \cite{BugniCanayShi2017} and \cite{KaidoMolinariStoye2019} in the simulations as additional points of comparison.

This paper is more closely related to the weak identification literature. 
We briefly discuss the connection to two main branches of the weak identification literature: estimation theory under weak identification and inference/hypothesis testing under weak identification. 

The literature on estimation theory under weak identification can be seen as a series of attempts to generalize the weak instruments model to cover nonlinear moments. 
All of the attempts require an assumption that classifies each parameter as strongly or weakly identified and that needs a reparameterization to be satisfied. 
The first attempt was by \cite{StockWright2000}, who derive limit theory for the generalized method of moments (GMM) estimator under weak identification. 
Assumption C in their paper requires a type of additive separability condition between the strongly and weakly identified parameters.
\cite{AndrewsCheng2012}, hereafter AC12, derive limit theory for a general class of extremum estimators under weak, semi-strong, or strong identification. 
Assumption A in that paper relaxes Assumption C in \cite{StockWright2000}. 
It only requires the objective function to be constant as a function of the weakly identified parameter when a subvector of the strongly identified parameters is equal to zero. 
\cite{AndrewsCheng2013,AndrewsCheng2014} verify the high-level conditions in AC12 for maximum likelihood, GMM, and nonlinear least squares objective functions, while \cite{Cheng2015} generalizes AC12 to allow for multiple sources of identification failure. 
\cite{HanMcCloskey2019} notice that a reparameterization satisfying Assumption A in AC12 is often difficult to find. 
They provide a systematic reparameterization procedure for satisfying the assumption.
In this paper, we propose a new structure that is suited to minimum distance models based on invertibility of the link function; see Assumption \ref{AssumptionNCIS}, below. 
The new structure is intuitive because it parallels the basic definition of identification as the invertibility of the parameterization of a model (the mapping from the parameters to the distribution of observables).

Existing weak identification estimation theory requires the parameter space to be a product space between the weakly identified parameters and the strongly identified parameters.\footnote{In \cite{StockWright2000}, see page 1060. In AC12, it follows from Assumption B1(ii), together with the definition of $\Pi$, that the parameter space must be a product space locally around $\beta=0$. The rest of the literature inherits the product space assumption from AC12.} 
This implies that the identified set is always the same (on the subset of the parameter space without identification); it is the set of all possible values for the weakly identified parameters. 
This precludes informative bounds on the identified set. 
(Trivial bounds that do not depend on estimated parameters are covered.) 
The product space assumption is especially concerning because of the reparameterization required to satisfy the weak identification structure. 
Even if one starts with an original parameterization that is a product space, it is unlikely that the new parameter space after reparameterization is a product space. 
This paper's solution is to relax the product space assumption. 
We derive new estimation limit theory that explicitly accounts for the influence of the bounds. 

This paper is also related to the literature on WIR inference/hypothesis testing. 
The goal of this literature is inference that is robust to the loss of identification in a uniform sense. 
This requires validity under non-identification (that is, when the true value of the parameter is not identified), as well as under weak identification (that is, when the true value of the parameter takes a sequence of values that converges to a point in the non-identified subset of the parameter space). 
\cite{StockWright2000}, \cite{Kleibergen2005}, \cite{GuggenbergerSmith2005}, \cite{Otsu2006}, \cite{Magnusson2010}, \cite{GuggenbergerRamalhoSmith2012}, \cite{Andrews2016}, \cite{AndrewsMikusheva2016a}, and \cite{AndrewsGuggenberger2019} provide tests for full-vector hypotheses or subvector hypotheses with strongly identified nuisance parameters. 
\cite{ChaudhuriZivot2011}, \cite{AndrewsMikusheva2016b}, and \cite{Andrews2017, Andrews2018} provide tests or confidence sets for subvector hypotheses with weakly identified nuisance parameters. 
\cite{Andrews2001} considers a special type of hypothesis testing problem when a parameter does not appear under the null hypothesis. 
AC12 recommend calculating a critical value by taking the supremum over quantiles of the limiting distributions under weak, semi-strong, or strong identification. 
The RQLR test is an adaptation of the approach in AC12 to the weak identification with bounds limit theory. 

It is also common in the WIR inference literature to require the parameter space to be a product space between the weakly and strongly identified parameters.\footnote{For example, see expression (3.1) in \cite{Andrews2001}, Assumption A in \cite{GuggenbergerSmith2005}, and (9.1) in \cite{AndrewsGuggenberger2019}.} 
When the parameter space is not a product space, the test may over-reject at the boundary of the identified set. 
This is because the argument that justifies plugging in strongly identified nuisance parameters requires the null-imposed estimator of the nuisance parameters to be asymptotically normal. 
When testing a hypothesis on the boundary of the identified set, the nuisance parameters are on the boundary of the parameter space. 
The resulting null-imposed estimator of the nuisance parameters is not asymptotically normal, invalidating the argument.\footnote{An exception is \cite{AndrewsGuggenberger2019}, which allows the nuisance parameters to be on the boundary but comes at the cost of power against hypotheses that violate the bounds.} 
These tests are discussed further in Section 5. 

This paper is also related to the literature that allows the parameter to be on or near the boundary of the parameter space; see \cite{Andrews1999}, \cite{MoonSchorfheide2009}, \cite{Ketz2018}, \cite{KetzMcCloskey2023}, and \cite{FanShi2023}. 
This literature requires strong identification. 

This paper is also related to the uniformity literature. 
\cite{AndrewsGuggenberger2010ET} and \cite{AndrewsChengGuggenberger2020} provide general results that characterize uniform asymptotic validity using limit theory under a comprehensive class of sequences of true parameter values. 
AC12 use this sequential characterization of uniformity to justify the consideration of three strengths of identification: weak, semi-strong, and strong. 
We follow this justification for considering weak, semi-strong, and strong identification.  
\cite{McCloskey2017} recommends a general approach to uniformity in models with a discontinuity in the limit theory based on a first-stage confidence set for local parameters. 
\cite{HanMcCloskey2019} use this approach in weakly identified models. 
The RQLR test also uses a first-stage confidence set for local parameters that enter the weak identification with bounds limit theory. 

This paper is also related to \cite{CoxWeakIdFactor}. 
\cite{CoxWeakIdFactor} provides reparameterizations for low-dimensional factor models with one or two factors to classify the parameters as strongly or weakly identified. 
The reparameterized models satisfy the identification structure required by Assumption \ref{AssumptionNCIS}, below. 
\cite{CoxWeakIdFactor} then compares WIR tests without bounds and finds that those tests that use the reparameterization to plug in strongly identified nuisance parameters are less conservative than tests that do not use the reparameterization. 
\cite{CoxWeakIdFactor} also shares an empirical application with this paper. 
\cite{CoxWeakIdFactor} points out the need for WIR inference and demonstrates the smaller confidence intervals that come from using the reparameterization to make available the less conservative WIR tests. 
In this paper, we demonstrate further gains that come from using bounds. 
In particular, one of the confidence intervals in \cite{CoxWeakIdFactor} is very wide and includes unreasonable values of the parameters. 
In this paper, we are able to exclude those values from the confidence interval. 

\textbf{Outline.} 
The rest of the paper is organized as follows. 
Section 2 defines the class of minimum distance models and presents the examples. 
Section 3 states the limit theory for the estimator. 
Section 4 states the limit theory for the QLR statistic and defines the RQLR test. 
Section 5 provides simulations of WIR tests in the two example factor models. 
Section 6 presents the empirical application. 
Section 7 concludes. 
Proofs and details for the examples and application are available in the supplemental materials. 

\textbf{Notation.} 
For clarity, we introduce some notation used throughout the paper. 
We write $(a,b)$ for $(a',b')'$ for stacking two column vectors, $a$ and $b$. 
We write $d_a$ to denote the dimension of an arbitrary vector, $a$. 
Let $e_1,...,e_d$ denote the standard normal basis vectors in $\R^d$. 
Given a function $f$, we write $f(a,b)$ for $f((a,b))$ for simplicity.  
Given a real-valued function $f(x)$, we write $\partial_x f(x)$ to denote the row vector of derivatives of $f(x)$. 
If $f(x)$ is a vector-valued function, $\partial_x f(x)$ denotes the matrix of derivatives where each row denotes the derivatives of the corresponding component of $f(x)$. 
If $f$ is real-valued, we use $\partial_{xy} f(x,y)$ and $\partial_{xx}f(x)$ to denote $\partial_y (\partial_x f(x,y))'$ and $\partial_x (\partial_x f(x))'$, respectively, which are matrices of second derivatives. 
Given any two subsets of a metric space, $A$ and $B$, let $\vec d(A,B)=\max(\sup_{a\in A}\inf_{b\in B} d(a,b),0)$ denote a directed distance between $A$ and $B$, where $d(\cdot,\cdot)$ is the metric. 
(The max with zero is taken so that $\vec{d}(A,B)=0$ when $A$ is the empty set.) 
We write the Hausdorff distance between two sets, $A$ and $B$, as $d_H(A,B)=\max\left(\vec d(A,B),\vec d(B,A)\right)$. 
Let $\rightarrow_H$ denote convergence in the Hausdorff distance. 
For a sequence of stochastic processes, $Q_n(\theta)$ and $Q(\theta)$ over $\theta\in K$, let $Q_n(\theta)\underset{K}{\Rightarrow}Q(\theta)$ denote weak convergence in the space of bounded functions on $K$ with the uniform norm. 

\section{A Class of Minimum Distance Models}

Minimum distance models are commonly used to estimate and test hypotheses on structural parameters of interest. 
This section introduces a class of minimum distance models that satisfy a key assumption and describes two example models that satisfy the assumption. 

In minimum distance models, the link function, $\delta(\cdot)$, maps structural parameters, $\theta$, to reduced-form parameters, $\delta$.\footnote{Abusing notation, $\delta$ denotes both the mapping and the reduced-form parameters themselves.} 
Reduced-form parameters are typically chosen to be features of the observed variables, including means, variances, covariances, or regression coefficients, that are informative for $\theta$. 
We assume the reduced-form parameters are identified, while identification of the structural parameters is determined by the link function. 
That is, a given parameter value $\theta^*$ is identified if $\delta(\theta^*)=\delta(\theta)$ implies $\theta^*=\theta$. 

Weak identification arises when $\theta$ is generically identified, but there is a problematic subset of the parameter space where $\theta$ is not identified. 
Let $\theta=(\pi,\beta)$, where $\pi$ denotes the structural parameters that are always identified and $\beta$ denotes the structural parameters that are possibly not identified. 
If $\pi$ is identified, then one can optimistically reparameterize $\theta$ so that $\pi$ are also reduced-form parameters. 
That is, $\delta=(\pi,\tau)$, where the components of $\pi$ are common to both $\theta$ and $\delta$, and $\tau$ are additional reduced-form parameters. 
This structure is helpful for analyzing the identification of $\beta$ and especially for defining weak identification using sequences of the identified parameters in $\pi$. 

The following assumption formalizes this structure. 
Let $\breve{\Theta}$ be an open set that denotes the parameter space for $\theta$. 
\begin{assumption}
\label{AssumptionNCIS} There exists a twice continuously differentiable function $\tau:\breve{\Theta}\rightarrow\R^{d_\tau}$ such that $\delta(\pi,\beta)=(\pi,\tau(\pi,\beta))$ for all $(\pi,\beta)\in\breve{\Theta}$. 
\end{assumption}
\textbf{Remarks:}
\textbf{2.1.} 
Assumption \ref{AssumptionNCIS} isolates $\tau(\pi,\beta)$ as a mapping that determines identification of $\beta$. 
For a fixed value of $\pi$, $\beta$ is identified if and only if $\tau(\pi,\beta)$ can be inverted for $\beta$. 
The set of $\pi$ values for which $\tau(\pi,\beta)$ cannot be inverted for $\beta$ defines the problematic subset of the parameter space. 
It is common in the weak identification literature to assume some structure that isolates the source of identification failure; see Assumption C in \cite{StockWright2000} or Assumption A in AC12. 
Assumption \ref{AssumptionNCIS} defines a type of identification structure for minimum distance models that parallels the basic definition of identification as the invertibility of the link function. 

\textbf{2.2.} The structure of identification required by Assumption \ref{AssumptionNCIS} is related to the structure required by Assumption A in AC12. 
If $\tau(\pi,\beta)=\pi_1\tilde\tau(\pi,\beta)$, where $\pi_1$ is a component of $\pi$ and $\tilde\tau(\pi,\beta)$ is a mapping that is invertible for $\beta$ for any fixed value of $\pi$, then the structure required by Assumption A in AC12 is satisfied. 
In this case, $\beta$ is identified if and only if $\pi_1\neq 0$. 

\textbf{2.3.} Assumption \ref{AssumptionNCIS} typically requires a reparameterization to ensure $\pi$ is common. 
While there is no guarantee for finding such a reparameterization, a general strategy is to replace a structural parameter with a reduced-form parameter if the link function for that parameter is invertible. 
If enough structural parameters are replaced with reduced-form parameters, then the remaining structural parameters are potentially weakly identified. 
In that case, the link function defines a reparameterization that satisfies Assumption \ref{AssumptionNCIS}. 
This strategy was used to find the reparameterizations for all the examples in this paper. 
\qed\medskip

Information about the structural parameters may also be available from bounds on the parameter space. 
Let $\ell: \breve{\Theta}\rightarrow \R^{d_\ell}$ define the bounds, where $d_\ell$ is the number of bounds. 
Each component of $\ell(\theta)$ specifies an inequality that restricts the possible values of $\theta$. 
We then define $\Theta=\{\theta\in\breve{\Theta}: \ell(\theta)\le 0\}$ to be the parameter space with boundary. 
Importantly, $\Theta$ is not required to be a product space between $\pi$ and $\beta$. 
The bounds could come from knowledge of the signs of one or more parameters. 
Also, moment inequalities that are additively separable can be written in this way by adding the expectations of the random parts of the moments as additional parameters. 

The bounds are informative because they rule out possible values of $\theta$. 
This is especially important without identification because then the identified set is smaller---it is restricted to values of $\theta$ in $\Theta$. 
If one were to ignore the bounds, then the identified set would consist of all observationally equivalent values of $\theta$ in $\breve{\Theta}$, a potentially much larger set. 

In the remainder of this section, we describe two example models and demonstrate how to reparameterize them to satisfy Assumption 1. 

\begin{namedexample}[1: Factor Model]
A factor model states that a $p$-vector of observed variables, $X_i$, is related to an $m$-vector of unobserved factors, $f_i$, by the equation, 
\begin{equation}
X_i=\Lambda f_i+\epsilon_i,\label{2.1}
\end{equation}
where $\epsilon_i$ is a $p$-vector of unobserved errors and $\Lambda$ is a $p\times m$ matrix of factor loadings.\footnote{For a classical treatment of factor models, see \cite{Gorsuch2014}.} 
Factor models are widely used, not only in economics, but throughout the social sciences. 
Identification is determined by the covariance matrix of $X_i$: 
\begin{equation}
Cov(X_i)=\Lambda\Sigma\Lambda'+\Phi, \label{2.2}
\end{equation}
where $\Sigma$ is an $m\times m$ symmetric and positive definite matrix that denotes the covariance matrix of the factors, $\Phi$ is a $p\times p$ diagonal matrix that denotes the covariance matrix of the errors, and $\epsilon_i$ is uncorrelated with $f_i$. 
The covariance matrix of $X_i$ is identified and can be written as a nonlinear function of the parameters. 
The factor model parameters are identified if and only if this nonlinear function can be inverted. 
\cite{AndersonRubin1956} give necessary and sufficient conditions on the factor loadings for identification. 
The conditions ensure enough loadings are nonzero and/or certain submatrices of the loading matrix are nonsingular.\footnote{Also note that conditions for identification in factor models do not depend on whether a ``high-dimensional'' or ``low-dimensional'' asymptotic approximation is used for the limit theory. Identification is a finite-sample property of the model.} 
These conditions are widely used in applications of factor models to ensure that, inter alia, first-stage estimation of the number of factors is asymptotically negligible. 
However, these conditions are rarely defended. 
Weak identification in factor models relaxes these conditions. 

Consider a factor model with only one factor and three observed variables. 
In this case, the factor loading matrix is given by 
\[
\Lambda=\left[\begin{array}{c}1\\\lambda_2\\\lambda_3\end{array}\right], 
\]
where $\lambda_1=1$ normalizes the scale of the factor in terms of $X_{1i}$. 
The other parameters in the model are the (scalar) variance of the factor, $\sigma^2$, and the variances of the errors, $\Phi=diag(\phi_1, \phi_2,\phi_3)$. 
We take the elements of $Cov(X_i)$ to be the reduced-form parameters. 
Writing out equation (\ref{2.2}), 
\begin{equation}
\left[\begin{array}{ccc}\omega_1&\rho_2&\rho_3\\\rho_2&\omega_2&\tau\\\rho_3&\tau&\omega_3\end{array}\right]=\left[\begin{array}{ccc}\sigma^2+\phi_1&\lambda_2\sigma^2&\lambda_3\sigma^2\\\lambda_2\sigma^2&\lambda_2^2\sigma^2+\phi_2&\lambda_2\lambda_3\sigma^2\\\lambda_3\sigma^2&\lambda_2\lambda_3\sigma^2&\lambda_3^2\sigma^2+\phi_3\end{array}\right], \label{2.3}
\end{equation}
where $\omega_j=Var(X_{ji})$, $\rho_j=Cov(X_{1i},X_{ji})$, and $\tau=Cov(X_{2i},X_{3i})$. 
While we only consider three observed variables, the generalization to more observed variables is straightforward using the reparameterizations in \cite{CoxWeakIdFactor}. 
The supplemental materials include an example factor model with two factors. 

We use equation (\ref{2.3}) to define a reparameterization: $(\lambda_2, \lambda_3, \sigma^2, \phi_1, \phi_2, \phi_3)\mapsto (\rho_2, \rho_3, \omega_1, \omega_2, \omega_3, \beta)$, where $\beta=\sigma^2$. 
\cite{CoxWeakIdFactor} proposes this reparameterization to analyze weak identification without bounds. 
The reparameterized model satisfies Assumption \ref{AssumptionNCIS} by setting $\pi=(\rho_2, \rho_3, \omega_1, \omega_2, \omega_3)$.\footnote{Note that the formula for the reparameterization comes from the link function in (\ref{2.3}), which demonstrates the general strategy for finding a reparameterization in Remark 2.3.} 
The function $\tau(\pi,\beta)$ is found by looking at the unused equation in (\ref{2.3}), associated with $\tau$, written in terms of the new parameters: 
\begin{equation}
\tau(\pi,\beta)=\rho_2\rho_3\beta^{-1}. 
\end{equation}
Thus, $\beta$ is not identified when either $\rho_2=0$ or $\rho_3=0$. 
This corresponds to $\lambda_2=0$ or $\lambda_3=0$. 
Thus, the variance of the factor is identified if and only if all three observed variables have nonzero factor loadings. 
In this paper, we assume $\rho_2\neq 0$ in order to focus on the effect of $\rho_3$. 

When $\rho_3=0$, $\beta$ is still partially identified by bounds. 
These bounds come from the nonnegativity of the variance parameters. 
The inequality $\phi_1\ge 0$, rewritten in the new parameters, is $\beta\le\omega_1$, which is an upper bound on $\beta$. 
Similarly, $\phi_2\ge 0$, rewritten in the new parameters, is $\beta\ge \rho_2^2/\omega_2$. 
Intuitively, the variance of the factor cannot be larger than the variance of the normalizing variable, and the variance of the factor must be large enough to explain the covariance between $X_{1i}$ and $X_{2i}$. 
Thus, the bounds can be written as 
\begin{equation}
\ell(\theta)=\left[\begin{array}{c}\beta-\omega_1\\\rho_2^2-\omega_2\beta\end{array}\right], \label{Ex1_ell_function}
\end{equation}
and therefore the identified set for $\beta$, when $\rho_3=0$, is $[\omega_2^{-1}\rho_2^2,\omega_1]$. 
As the variances of the errors get smaller, so that the factors are measured with less error, the interval shrinks. 
Under weak identification, these bounds provide a valuable source of information about $\beta$. 
\qed  
\end{namedexample}

\begin{namedexample}[2: GARCH Model] 
Models of autoregressive conditional heteroskedasticity are often used for financial and economic time series. 
Let $Y_t$ denote a strictly stationary time series for $t\in\{1,...,n\}$. 
A simple GARCH(1,1) model is given by 
\begin{align}
Y_t&=\mu+\epsilon_t\nonumber\\
\sigma^2_t&=(1-\alpha-\beta)\sigma^2+\beta \sigma^2_{t-1}+\alpha\epsilon^2_{t-1}, \label{GARCH_equation}
\end{align}
where $\epsilon_t$ is a martingale difference sequence with $\sigma^2_t=\mathbb{E}[\epsilon_t^2|Y_1,...,Y_{t-1}]$ the conditional variance. 
In (\ref{GARCH_equation}), $\mu$, $\sigma^2$, $\alpha$, and $\beta$ are unknown parameters, where $\mu$ is the mean of $Y_t$ and $\sigma^2$ is the variance of $Y_t$. 
The conditional mean of $Y_t$ is modeled as simply as possible with just a constant. 
(Generalizations to allow for a conditional mean function that depends on lags of $Y_t$, $\epsilon_t$, or other variables in the dataset are straightforward.) 
The ``(1,1)'' indicates that one lag of $\sigma^2_t$ and one lag of $\epsilon^2_t$ are used to model the conditional variance. 
This model was first proposed by \cite{Bollerslev1986}; see \cite{FrancqZakoian2019} for a textbook treatment. 

To ensure the conditional variance is nonnegative, some parameter restrictions are added: $\sigma^2>0$, $\alpha\ge 0$, and $\beta\ge 0$. 
In addition, to ensure the conditional variance is stationary, it is typically assumed that $\alpha+\beta<1$. 
Here, we strengthen that assumption to $\alpha+\beta\le 1-\eta$ for some $\eta>0$ in order to bound the conditional variance process away from nonstationarity. 

While a GARCH model is usually estimated by (quasi) maximum likelihood, it can be cast as a minimum distance model with reduced-form parameters given by the autocorrelations of $\epsilon_t^2$.\footnote{We can estimate $\mu$ and $\sigma^2$ with the sample mean and variance  of $Y_t$, so the only parameters that need to be estimated by minimum distance are $\alpha$ and $\beta$. More generally, the autocorrelations of $Y_t$ can be used as reduced-form parameters for a model of the conditional mean of $Y_t$.} 
One can write the first $k$ autocorrelations of $\epsilon_t^2$ in terms of the model parameters: 
\begin{align}
\text{corr}(\epsilon_t^2, \epsilon_{t-1}^2)&=\alpha\frac{1-\alpha\beta-\beta^2}{1-2\alpha\beta-\beta^2}\nonumber\\
\text{corr}(\epsilon_t^2, \epsilon_{t-2}^2)&=\alpha\frac{1-\alpha\beta-\beta^2}{1-2\alpha\beta-\beta^2}(\alpha+\beta)\nonumber\\
\vdots~&=~\vdots\nonumber\\
\text{corr}(\epsilon_t^2, \epsilon_{t-k}^2)&=\alpha\frac{1-\alpha\beta-\beta^2}{1-2\alpha\beta-\beta^2}(\alpha+\beta)^{k-1}.  \label{GARCH_autocorrrelations}
\end{align}
When $\alpha=0$, these autocorrelations are all zero and the value of $\beta$ cannot be identified. 
This indicates that the model is weakly identified when $\alpha$ is near zero. 

We can reparameterize the model to satisfy Assumption \ref{AssumptionNCIS}. 
Let 
\[
\pi=\alpha\frac{1-\alpha\beta-\beta^2}{1-2\alpha\beta-\beta^2} 
\]
be a new parameter defined as a function of $\alpha$ and $\beta$. 
(Note that $\pi=\text{corr}(\epsilon_t^2, \epsilon_{t-1}^2)$, following the general reparameterization strategy from Remark 2.3.) 
For any value of $\beta$, this can be inverted for a unique value of $\alpha$. 
Let $\alpha(\pi,\beta)$ denote the inverse.\footnote{It is possible to solve for the inverse using the quadratic formula: 
\[
\alpha(\pi,\beta) = \begin{cases}
\pi+\frac{1-\beta^2}{2\beta}-\sqrt{\pi^2+\left(\frac{1-\beta^2}{2\beta}\right)^2}& \text{ if }\beta>0\\
\pi & \text{ if }\beta=0
\end{cases}. 
\]}
If we let $\tau_j=\text{corr}(\epsilon_t^2,\epsilon_{t-j}^2)$ for $j\in\{2,...,k\}$, then the reparameterized model satisfies Assumption \ref{AssumptionNCIS} with $\tau=(\tau_2,...,\tau_k)$, $\theta=(\pi,\beta)$, and $\delta=(\pi,\tau)$. 
Note that $\tau_j(\pi,\beta)=\pi(\alpha(\pi,\beta)+\beta)^{j-1}$ gives the reparameterized link function. 
When $\pi=0$ (which is equivalent to $\alpha=0$), the link function cannot be inverted for $\beta$. 

When $\pi=0$, bounds on the identified set for $\beta$ are given by rewriting the above inequalities. 
The lower bound is given by $\beta\ge 0$, while the upper bound is given by $\alpha(\pi,\beta)+\beta\le 1-\eta$. 
We can write these bounds as 
\begin{equation}
\ell(\theta)=\left[\begin{array}{c}-\beta\\\alpha(\pi,\beta)+\beta+\eta-1\end{array}\right]. \label{GARCH_ell_definition}
\end{equation}
Note that the parameter space is not a product space between $\beta$ and $\pi$. 
However, when $\pi=0$, the identified set for $\beta$ simplifies to $[0,1-\eta]$. 
\qed
\end{namedexample}

Any minimum distance model that combines weak identification with bounds is a potential application. 
Section A, in the Supplemental Materials, gives six other examples, including a factor model with two factors, a linear instrumental variables model, an autoregressive moving average model, a threshold crossing model, a model of household expenditures, and a structural vector autoregression. 

\section{Estimation Limit Theory}

This section states conditions for limit theory of a minimum distance estimator under weak identification with bounds. 
We fix a sequence of parameter values that is indexed by the sample size and determines the strength of identification. 
Let $P$ denote the distribution of a sample, $W_1,...,W_n$, indexed by the sample size, $n$. 
Let $P=P_{\theta,\xi}$ depend on the structural parameters, $\theta$, and an additional, possibly infinite dimensional parameter, $\xi$. 
We fix a sequence of true parameter values, $\theta_n=(\pi_n,\beta_n)\rightarrow\theta_\ast=(\pi_\ast,\beta_\ast)$ and $\xi_n\rightarrow\xi_\ast$, indexed by the sample size. 
Throughout this section, limits are taken as $n\rightarrow\infty$ and probabilities are calculated using $P=P_{\theta_n,\xi_n}$. 

The overall strategy is to start with assumptions on the reduced-form model. 
Since the reduced-form parameters are identified, familiar assumptions from the literature can be used to ensure regular limit theory. 
That limit theory is then mapped to limit theory for the estimator of the structural parameters depending on whether $\tau(\pi_\ast,\beta)$ is invertible for $\beta$. 

The assumptions must also be carefully formulated to account for the bounds by allowing the sequence of parameters to be on or near the boundary of the parameter space. 
In particular, the conditions avoid the assumption that the parameter space is a product space between the strongly identified parameters and the weakly identified parameters. 
Subsection \ref{Section_3.4} uses a setwise limit of the local parameter space to account for the influence of the boundary. 

\subsection{Consistency}

This subsection defines the minimum distance estimator and states sufficient conditions for a type of consistency result. 
Let $Q_n(\theta)$ denote a real-valued data-dependent objective function on $\breve{\Theta}$. 
We take the estimator, $\hat\theta_n$, to be a random vector in $\Theta$ satisfying 
\begin{equation}
Q_n(\hat{\theta}_n)\le \inf_{\theta\in\Theta} Q_n(\theta)+o_p(n^{-1}).\label{MEstimator}
\end{equation}
We write $\hat\theta_n=(\hat{\pi}_n,\hat{\beta}_n)$. 
This definition of $\hat{\theta}_n$ imposes the bounds through the boundary of $\Theta$. 

The following assumption says that $Q_n(\theta)$ depends on $\theta$ only through the reduced-form parameters, $\delta$. 
\begin{assumption}
\label{AssumptionRFO}
There exists a real-valued data-dependent function $T_n(\delta)$ on $\delta(\breve{\Theta})$ that is almost surely twice differentiable and satisfies $T_n(\delta(\theta))=Q_n(\theta)$ for all $\theta\in\breve{\Theta}$. 
\end{assumption}
\noindent\textbf{Remark:} 
\textbf{3.1.} Assumption \ref{AssumptionRFO} is the defining feature of a minimum distance objective function. 
It is satisfied in Examples 1 and 2. 
\qed\medskip

Because of Assumption \ref{AssumptionRFO}, we can view $T_n(\delta)$ as an objective function that defines an extremum estimator of identified parameters and place familiar assumptions from the extremum estimation literature to get regular limit theory. 
The following assumption ensures the reduced-form parameters are identified and $T_n(\delta)$ can consistently estimate them. 
\begin{assumption}
\label{AssumptionRFI} There exists a nonstochastic real-valued function, $T(\delta)$, continuous on $\delta(\breve{\Theta})$, such that 
\begin{enumerate}[label=(\alph*)]
\item $\sup_{\delta\in\delta(\breve{\Theta})}|T_n(\delta)-T(\delta)|\rightarrow_p 0$, and 
\item for every neighborhood, $U$, of $\delta_\ast=\delta(\theta_\ast)$,  
\begin{equation*}
\inf_{\delta\in \delta(\breve{\Theta})/U} T(\delta)-T(\delta_\ast)>0.
\end{equation*}
\end{enumerate}
\end{assumption}

Assumption \ref{AssumptionRFI} is sufficient for consistent estimation of the reduced form parameters, including $\pi$. 
However, when the parameter space is not a product space, we also need to show that the support of the distribution of $\hat{\beta}_n$ concentrates on the identified set. 
For this purpose, the next assumption places a restriction on the shape of the parameter space. 
For any $\Pi\subseteq \R^{d_\pi}$, let 
\begin{align*}
\mathcal{B}(\Pi)&=\{\beta\in\R^{d_\beta}| (\pi,\beta)\in\Theta \text{ for some }\pi\in\Pi\},  
\end{align*}
which defines the cross-section of the parameter space. 
Let $\mathcal{B}_\ast=\mathcal{B}(\pi_\ast)$ denote the identified set for $\beta$ when $\tau(\pi_\ast,\beta)$ is not invertible for $\beta$. 
Recall $d_H(A,B)$ denotes the Hausdorff distance between sets $A$ and $B$. 

\begin{assumption}
\label{AssumptionPSS}
\begin{enumerate}[label=(\alph*)]
\item $\Theta$ is closed, $\mathcal{B}_\ast$ is compact, and 
\item there exists a neighborhood of $\pi_\ast$ and an $M>0$ such that for all $\pi_1$ and $\pi_2$ in the neighborhood, $$d_H(\mathcal{B}(\pi_1),\mathcal{B}(\pi_2))\le M \|\pi_1-\pi_2\|.$$
\end{enumerate}
\end{assumption}

\textbf{Remark:} 
\textbf{3.2.} Part (b) assumes the cross-section for $\beta$ is locally Lipschitz continuous in the Hausdorff distance as a function of $\pi$. 
Importantly, this does not require differentiability of the bounds, thus permitting kink points associated with the intersection of multiple bounds. 
When $\mathcal{B}(\pi)$ is an interval, as in Examples 1 and 2, Assumption \ref{AssumptionPSS}(b) only requires the endpoints of the interval to be locally Lipschitz continuous in $\pi$. 
\qed 

\begin{lemma}
\label{LemmaCON}
Under Assumptions 1, 2, 3, and 4, there exists a sequence of neighborhoods of $\pi_\ast$, $\Pi_n$, such that $\mathcal{B}_n:=\mathcal{B}(\Pi_n)$ is compact and satisfies 
\begin{enumerate}[label=(\alph*)]
\item $\Pi_n\supseteq\Pi_{n+1}$, $\mathcal{B}_n\supseteq\mathcal{B}_{n+1}$, and $\Pi_n\times \mathcal{B}_n\subseteq\breve{\Theta}$ for all $n$, $\Pi_n\rightarrow_H\{\pi_\ast\}$, $\mathcal{B}_n\rightarrow_H\mathcal{B}_\ast$, and 
\item $P(\hat\theta_n\in\Pi_n\times\mathcal{B}_n)\rightarrow 1$. 
\end{enumerate}
\end{lemma}

\textbf{Remark:} 
\textbf{3.3.} Lemma \ref{LemmaCON} shows that $\hat{\pi}_n\rightarrow_p\pi_\ast$ and that the support of the distribution of $\hat{\beta}_n$ concentrates on $\mathcal{B}_\ast$. 
Here, $\Pi_n\times \mathcal{B}_n$ denotes a sequence of neighborhoods of the identified set that is converging monotonically downward to the identified set. 
The use of $\Pi_n\times \mathcal{B}_n$ in the derivation of the limit theory is a novel proof strategy that is needed to handle informative bounds. 
\qed 

\begin{namedexample}[1, Continued]
We estimate Example 1 using a minimum distance objective function that is quadratic in the reduced-form parameters. 
Let $\widehat\Omega$ denote an estimator of the covariance matrix of $X_i$, and let 
\begin{equation}
Q_n(\theta)=(\delta(\theta)-vech(\widehat\Omega))'\widehat W_n(\delta(\theta)-vech(\widehat\Omega))/2, 
\end{equation}
denote the objective function, where $\widehat W_n$ converges in probability to a positive definite weight matrix.\footnote{The $vech(\cdot)$ function vectorizes symmetric matrices. We take the ordering of the elements in $vech$ to coincide with $\delta$.} 

The limit theory is driven by the limit theory of $\widehat\Omega$. 
We assume $\widehat\Omega$ is consistent and asymptotically normal along the sequence $P_{\theta_n,\xi_n}$. 
That is, there exists a $d_\delta\times d_\delta$ positive definite matrix, $\bar V$, such that $\sqrt{n}(vech(\widehat\Omega)-\delta(\theta_n))\rightarrow_d N(0,\bar V)$. 
This can be verified by a triangular-array central limit theorem, for example if $X_i$ is independent across $i$ with $4+\epsilon$ finite moments. 
Alternative central limit theorems that allow dependence are also possible. 
Let $\hat{\bar V}$ be a consistent estimator of $\bar V$. 
If $\widehat W_n=\hat{\bar V}\inv$, then the objective function is optimally weighted, but this is not required. 
\end{namedexample}

\begin{namedexample}[2, Continued]
Estimating a GARCH model by minimum distance is possible by defining the reduced-form parameters to be the autocorrelations of $\epsilon_t^2$. 
Let $\hat\delta$ denote a vector of estimators of the first $k$ autocorrelations of $\epsilon_t^2$. 
For example, for the $j^{\text{th}}$ autocorrelation, we can take $\hat\delta_j=\hat\gamma_j/\hat\gamma_0$, where $\hat\gamma_j=n^{-1}\sum_{t=j+1}^n (\hat{\epsilon}_t-\hat\sigma^2)(\hat{\epsilon}_{t-j}-\hat\sigma^2)$, $\hat\epsilon_t=Y_t-\bar Y$, $\hat\sigma^2=n^{-1}\sum_{t=1}^n\hat{\epsilon}_t^2$, and $\bar Y=n^{-1}\sum_{t=1}^n Y_t$. 
We can take the minimum distance objective function to be quadratic in the reduced-form parameters: 
\begin{equation}
Q_n(\theta)=(\delta(\theta)-\hat\delta)'\widehat W_n(\delta(\theta)-\hat\delta)/2, 
\end{equation}
where $\widehat W_n$ converges in probability to a positive definite weight matrix. 
Under standard conditions, $\hat\delta$ is asymptotically normal with some asymptotic variance matrix, $\bar V$. 
Let $\hat{\bar{V}}$ be a consistent estimator of $\bar V$. 
If $\widehat W_n=\hat{\bar{V}}^{-1}$, then the objective function is optimally weighted, but this is not required. 
\qed
\end{namedexample}

\subsection{Identification Strength}

The limit theory depends on the strength of identification. 
There are three identification strengths, strong, semi-strong, and weak, determined by the sequence of true values of the parameters. 
The relevant quantity for determining identification strength is $\tau(\pi_n,\beta)-\tau(\pi_n,\beta_n)$, viewed as a function of $\beta$. 
The next assumption defines different types of sequences based on whether $\tau(\pi_n,\beta)-\tau(\pi_n,\beta_n)$ converges to zero, and, if so, the rate of convergence. 
Let $\partial_{\pi}\tau(\pi,\beta)$ and $\partial_\beta \tau(\pi,\beta)$ denote the $d_\tau\times d_\pi$ and $d_\tau\times d_\beta$ matrices of derivatives of $\tau(\pi,\beta)$ with respect to $\pi$ and $\beta$, respectively. 

\begin{namedassumption}[5(S)]
For a given sequence $\{\theta_n\}_{n=1}^\infty$ such that $\theta_n\rightarrow \theta_\ast$, 
\begin{enumerate}[label=(\alph*)]
\item $\tau(\pi_\ast,\beta)=\tau(\pi_\ast,\beta_\ast)$ implies $\beta=\beta_\ast$, and 
\item $C:=\partial_\beta \tau(\pi_\ast,\beta_\ast)$ has full rank $d_\beta$. 
\end{enumerate}
\end{namedassumption}

\begin{namedassumption}[5(SS)]
For a given sequence $\{\theta_n\}_{n=1}^\infty$ such that $\theta_n\rightarrow \theta_\ast$, there exists a sequence of positive constants, $a_n$, such that $a_n\rightarrow \infty$ and $n^{-1/2}a_n\rightarrow 0$ as $n\rightarrow\infty$. 
\begin{enumerate}[label=(\alph*)]
\item There exists a function, $c(\beta)$, such that 
\[
a_n(\tau(\pi_n,\beta)-\tau(\pi_n,\beta_n))\rightarrow c(\beta),
\]
uniformly over $\beta\in\mathcal{B}_1$, and $c(\beta)=0$ implies $\beta=\beta_\ast$.\footnote{Here and throughout, $\mathcal{B}_1$ denotes the first element of the sequence $\mathcal{B}_n$.} 
\item There exists a matrix-valued function, $C(\beta)$, such that 
\[
a_n\partial_\beta \tau(\pi_n,\beta)\rightarrow C(\beta),
\]
uniformly over $\beta$ in a neighborhood of $\beta_\ast$, and $C:=C(\beta_\ast)$ has full rank $d_\beta$. 
\end{enumerate}
\end{namedassumption}

\begin{namedassumption}[5(W)]
For a given sequence $\{\theta_n\}_{n=1}^\infty$ such that $\theta_n\rightarrow \theta_\ast$, there exists a function, $c(\beta)$, such that 
\[
\sqrt{n}(\tau(\pi_n,\beta)-\tau(\pi_n,\beta_n))\rightarrow c(\beta),
\]
uniformly over $\beta\in \mathcal{B}_1$. 
\end{namedassumption}
\stepcounter{assumption}

\textbf{Remarks:} 
\textbf{3.4.} Assumption 5 characterizes different types of sequences according to the rate $\tau(\pi_n,\beta)-\tau(\pi_n,\beta_n)$ converges to zero as a function of $\beta\in\mathcal{B}_1$. 
``S'' stands for strongly identified, ``SS'' stands for semi-strongly identified, and ``W'' stands for weakly identified. 
Assumption 5(S) is satisfied when $\tau(\pi_\ast,\beta)$ is injective as a function of $\beta$. 
When not injective, $\tau(\pi_n,\beta)-\tau(\pi_n,\beta_n)$ converges to zero. 
Assumption 5(SS) is satisfied when $\tau(\pi_n,\beta)-\tau(\pi_n,\beta_n)$ converges to zero slower than a $\sqrt{n}$ rate, while Assumption 5(W) is satisfied when this quantity converges to zero at a $\sqrt{n}$ rate or faster. 
Assumption 5(W) covers the important case that $\tau(\pi_n,\beta)$ is not injective as a function of $\beta$ for any $n$. 
This corresponds to $\pi_n=\pi_\ast$ for all $n$. 
In this case, $c(\beta)\equiv 0$. 

\textbf{3.5.} If $\tau(\pi,\beta)=\pi_1\tilde\tau(\pi,\beta)$, so Assumption A in AC12 is satisfied (see Remark 2.2), then the strength of identification is determined by the rate at which $\pi_1$ converges to zero. 
Weak identification occurs when $\pi_1$ converges to zero at an $n^{-1/2}$ rate or faster. 
If $\partial_\beta \tilde\tau(\pi_\ast,\beta_\ast)$ has full-rank, then semi-strong identification occurs when $\pi_1$ converges to zero slower than $n^{-1/2}$, and strong identification occurs when $\pi_1$ converges to a nonzero value. 
This agrees with the definitions of weak, semi-strong, and strong identification in AC12. 
\qed \medskip

In Assumption 5(SS), $a_n$ indexes the weakness of the sequence. 
By extension, we can take $a_n=1$ for all $n$ under Assumption 5(S) and $a_n=\sqrt{n}$ for all $n$ under Assumption 5(W). 

\begin{namedexample}[1, Continued]
In Example 1, $\tau(\pi,\beta)=\rho_2\rho_3/\beta$. 
Invertibility of this mapping for $\beta$ depends on the value of $\rho_3$. 
Assumption 5 then characterizes sequences of $\theta_n=(\rho_{2n},\rho_{3n},\omega_{1n},\omega_{2n},\omega_{3n},\beta_n)\rightarrow\theta_\ast=(\rho_{2\ast},\rho_{3\ast},\omega_{1\ast},\omega_{2\ast},\omega_{3\ast},\beta_\ast)$ based on the value of $\rho_{3n}$. 
Strong sequences have $\rho_{3\ast}\neq 0$. 
Weak sequences satisfy $\sqrt{n}\rho_{3n}\rightarrow s_\ast\in\R$. 
These sequences satisfy Assumption 5(W) with $c(\beta)=s_\ast\rho_{2\ast}(\beta\inv-\beta_\ast\inv)$. 
Semi-strong sequences have $\rho_{3\ast}=0$ and satisfy $\sqrt{n}\rho_{3n}\rightarrow \pm\infty$. 
These sequences satisfy Assumption 5(SS) with $c(\beta)=\pm\rho_{2\ast}(\beta\inv-\beta_\ast\inv)$, $C(\beta)=\mp\rho_{2\ast}\beta^{-2}$, and $a_n=|\rho_{3n}|\inv$. 
\qed
\end{namedexample}

\begin{namedexample}[2, Continued]
In Example 2, $\tau_j(\pi,\beta)=\pi(\alpha(\pi,\beta)+\beta)^{j-1}$ for $j\in\{2,...,k\}$. 
Invertibility of this mapping for $\beta$ depends on the value of $\pi$. 
Assumption 5 then characterizes sequences of $\theta_n=(\pi_n, \beta_n)\rightarrow \theta_\ast=(\pi_\ast, \beta_\ast)$ based on the value of $\pi_n$. 
Strong sequences have $\pi_\ast>0$. 
Weak sequences satisfy $\sqrt{n}\pi_n\rightarrow s_\ast\ge 0$. 
These sequences satisfy Assumption 5(W) with $c(\beta)=s_\ast(\beta^{j-1}-\beta^{j-1}_\ast)$. 
Semi-strong sequences have $\pi_\ast=0$ and satisfy $\sqrt{n}\pi_n\rightarrow\infty$. 
These sequences satisfy Assumption 5(SS) with $c(\beta)=\beta^{j-1}-\beta^{j-1}_\ast$, $C(\beta)=(j-1)\beta^{j-2}$, and $a_n=\pi_n^{-1}$. 
\qed
\end{namedexample}

\subsection{Convergence Rate}\label{Section_3.3}
This section derives the convergence rate of the components of $\hat\theta_n$ for the identification strengths defined in Section 3.2. 
We exploit the fact that the reduced-form model is strongly identified so we can use familiar regularity conditions. 
We first assume the reduced-form objective function is regular in the sense that it satisfies a quadratic expansion. 
Let $\delta_n=\delta(\theta_n)$. 
For every $\delta\in\delta(\breve{\Theta})$, let 
\begin{equation}\label{eq:QE}
R_n(\delta)=T_n(\delta)-T_n(\delta_n)-\partial_{\delta}T_n(\delta_n)'(\delta-\delta_n)-\frac{1}{2}(\delta-\delta_n)'\partial_{\delta\delta}T_n(\delta_n)(\delta-\delta_n),
\end{equation}
where $\partial_{\delta}T_n(\delta)$ denotes the first derivative vector and $\partial_{\delta\delta}T_n(\delta)$ denotes the second derivative matrix of $T_n(\delta)$. 
$R_n(\delta)$ is the remainder term in a quadratic expansion of $T_n(\delta)$ around $\delta=\delta_n$. 
\begin{assumption}
\label{AssumptionQE}
For a given sequence $\{\theta_n\}_{n=1}^{\infty}$ such that $\theta_n\rightarrow\theta_\ast$ and for every sequence of positive constants, $\eta_n\rightarrow 0$, 
\begin{equation*}
\sup_{\delta\in \delta(\breve{\Theta}):\|\delta-\delta_n\|\le \eta_n} \frac{|n R_n(\delta)|}{(1+\sqrt{n}\|(\delta-\delta_n)\|)^2}=o_p(1), 
\end{equation*}
where $\delta_n=\delta(\theta_n)$. 
\end{assumption}

In addition, we assume the derivatives of $T_n(\delta)$ satisfy regular limits so the reduced-form model is locally asymptotically normal around $\delta_n$. 
\begin{assumption}
\label{AssumptionRFL} 
For a given sequence $\{\theta_n\}_{n=1}^{\infty}$ such that $\theta_n\rightarrow\theta_\ast$: 
\begin{enumerate}[label=(\alph*)]
\item There exists a random vector, $Y$, and a positive definite and symmetric $d_\delta\times d_\delta$ matrix, $V$, such that 
\[
\sqrt{n}\partial_\delta T_n(\delta(\theta_n))\rightarrow_d Y\sim N(0,V).
\]
\item There exists a positive definite and symmetric $d_\delta\times d_\delta$ matrix, $H$, such that $\partial_{\delta\delta} T_n(\delta(\theta_n))\rightarrow_p H$. 
\end{enumerate}
\end{assumption}
\textbf{Remarks:} 
\textbf{3.6.} Assumption \ref{AssumptionQE} ensures the reduced-form objective function satisfies a quadratic expansion around $\delta_n=\delta(\theta_n)$. 
It is a type of stochastic differentiability condition as found in, for example, \cite{PakesPollard1989} or Section 3.2.4 in \cite{VaartWellner1996}. 
In Examples 1 and 2, Assumption \ref{AssumptionQE} is satisfied without remainder because $T_n(\delta)$ is quadratic in $\delta$. 
Examples where $T_n(\delta)$ is not quadratic are also covered, such as the maximum likelihood estimator of a factor model. 

\textbf{3.7.} Assumption \ref{AssumptionRFL}(a) can be verified by a central limit theorem, and Assumption \ref{AssumptionRFL}(b) can be verified by a law of large numbers. 
In Examples 1 and 2, if $W$ denotes the probability limit of $\widehat W_n$, then Assumption \ref{AssumptionRFL} is satisfied with $V=W \bar V W$ and $H=W$. 
When $W=\bar V\inv$, so the objective function is optimally weighted, $H=V=\bar V\inv$. 

\textbf{3.8.} These assumptions are reasonable for the reduced-form parameters because they are strongly identified. 
Analogous conditions on the structural parameters would need to be modified to account for the weak identification. 
See Assumption C1 in AC12 for the type of quadratic expansion that would be needed on the structural parameters. 
\qed\medskip

Under these assumptions, the following lemma states the rate of convergence of $\hat\theta_n$ for each identification strength. 
\begin{lemma}\label{Lemma2}
Under Assumptions \ref{AssumptionNCIS}, \ref{AssumptionRFO}, \ref{AssumptionRFI}, \ref{AssumptionPSS}, \ref{AssumptionQE}, and \ref{AssumptionRFL}, the following hold. 
\begin{enumerate}[label=(\alph*)]
\item Under Assumption 5(S), $\sqrt{n}(\hat\theta_n-\theta_n)=O_p(1)$. 
\item Under Assumption 5(SS), $\sqrt{n}(\hat{\pi}_n-\pi_n)=O_p(1)$ and $\sqrt{n}(\hat{\beta}_n-\beta_n)/a_n=O_p(1)$. 
\item Under Assumption 5(W), $\sqrt{n}(\hat{\pi}_n-\pi_n)=O_p(1)$ and $\hat{\beta}_n=O_p(1)$. 
\end{enumerate}
\end{lemma}

\textbf{Remark:} 
\textbf{3.9.} Lemma \ref{Lemma2} translates the rate of convergence from a regular reduced-form model to an irregular model for the structural parameters. 
In part (a), the rate is what we expect under strong identification. 
In part (c), under weak identification, $\hat{\beta}_n$ is no longer consistent. 
In part (b), under Assumption 5(SS), we see that the rate depends on how weak identification is. 
As identification gets weaker ($a_n$ diverges faster) the rate of convergence gets slower until $\hat{\beta}_n$ is not consistent. 
\qed \medskip

With the rate of convergence established by Lemma \ref{Lemma2}, the next step is to restandardize the objective function locally at the rate of convergence. 
Let $\psi=(\psi_1,\psi_2)\in\R^{d_\theta}$ denote local parameters, where $\psi_1\in\R^{d_\pi}$ and $\psi_2\in\R^{d_\beta}$. 
We define the local objective functions and their limits next. 
Let 
\begin{align*}
q^S_n(\psi)&=Q_n(\theta_n+n^{-1/2}\psi)-Q_n(\theta_n)\\
q^{SS}_n(\psi_1,\psi_2)&=Q_n(\pi_n+n^{-1/2}\psi_1,\beta_n+a_n n^{-1/2}\psi_2)-Q_n(\theta_n)\\
q^W_n(\psi_1,\beta)&=Q_n(\pi_n+n^{-1/2}\psi_1,\beta)-Q_n(\theta_n)
\end{align*}
denote the local objective functions. 
We use the following notation to define the limiting objective functions. Let 
\begin{equation}
D(\theta)=\partial_\theta \delta(\theta)=\left[\begin{array}{cc}I_{d_\pi}&0_{d_\pi\times d_\beta}\\\partial_\pi \tau(\theta)&\partial_\beta \tau(\theta)\end{array}\right]\label{D_definition}
\end{equation}
denote the derivative of the link function. 
Let $D_\ast=D(\theta_\ast)$. 
The limit of the local objective function under strong identification depends on $J=D'_\ast HD_\ast$ and $Z=-J\inv D'_\ast Y$.

The limit of the local objective function under semi-strong identification is very similar to the limit under strong identification. 
It depends on 
\begin{equation}
D_\ast = \left[\begin{array}{cc}I_{d_\pi}&0_{d_\pi\times d_\beta}\\\partial_\pi \tau(\theta_*)&C\end{array}\right], \label{D_definition_SS}
\end{equation}
where $C$ is defined in Assumption 5(SS). 
To cut down on notation, we write the strong and semi-strong limits using the same symbols, $J=D'_\ast HD_\ast$ and $Z=-J\inv D'_\ast Y$, with the implicit understanding that the definition of $D_\ast$ is different under semi-strong identification. 

The limit of the local objective function under weak identification depends on 
\begin{align}
D_1(\theta)&=\partial_\pi \delta(\theta)=\left[\begin{array}{c}I_{d_\pi}\\\partial_\pi \tau(\theta)\end{array}\right]\nonumber\\
J_{11}(\beta)&=D_1(\pi_\ast,\beta)'HD_1(\pi_\ast,\beta)\nonumber\\
g(\beta)&=\left[\begin{array}{c}0_{d_\pi\times 1}\\c(\beta)\end{array}\right]\nonumber\\
Z_1(\beta)&=-J_{11}\inv(\beta)D_1(\pi_\ast,\beta)'(Y+Hg(\beta))\nonumber\\
\tilde{q}^W(\beta)&=g(\beta)'Hg(\beta)/2+Y'g(\beta)-Z_1(\beta)'J_{11}(\beta)Z_1(\beta)/2,\label{3.7}
\end{align}
viewed as functions of $\beta\in\mathcal{B}_1$. 
We can then define 
\begin{align*}
q^S(\psi)&=(Z-\psi)'J(Z-\psi)/2-Z'JZ/2\\
q^W(\psi_1,\beta)&=(Z_1(\beta)-\psi_1)'J_{11}(\beta)(Z_1(\beta)-\psi_1)/2+\tilde{q}^W(\beta)
\end{align*}
to be the limits of the local objective functions under strong/semi-strong and weak identification. 
The following lemma shows weak convergence of the local objective functions uniformly over all compact subsets of their domains. 
(Recall $\underset{K}{\Rightarrow}$ denotes weak convergence in the space of bounded functions on $K$ with the uniform norm.) 
\begin{lemma}\label{Lemma3}
Under Assumptions \ref{AssumptionNCIS}, \ref{AssumptionRFO}, \ref{AssumptionRFI}, \ref{AssumptionPSS}, \ref{AssumptionQE}, and \ref{AssumptionRFL}, the following hold. 
\begin{enumerate}[label=(\alph*)]
\item Under Assumption 5(S), for all compact $K\subseteq\R^{d_\theta}$, $nq_n^S(\psi)\underset{K}{\Rightarrow} q^S(\psi)$. 
\item Under Assumption 5(SS), for all compact $K\subseteq\R^{d_\theta}$, $nq_n^{SS}(\psi)\underset{K}{\Rightarrow} q^S(\psi)$. 
\item Under Assumption 5(W), for all compact $K\subseteq\R^{d_\pi}\times\mathcal{B}_1$, $nq_n^W\hspace{-0.3mm}(\psi_1,\beta)\hspace{-0.2mm}\underset{K}{\Rightarrow} \hspace{-0.2mm}q^W\hspace{-0.3mm}(\psi_1,\beta)$. 
\end{enumerate}
\end{lemma}
\textbf{Remark:} 
\textbf{3.10.} The limit under strong and semi-strong identification is a quadratic function of $\psi$. 
Under weak identification, the limit is a quadratic function of $\psi_1$, but the coefficients in the quadratic function depend on $\beta$ in a non-quadratic way. 
This is the source of the nonstandard distribution under weak identification. \qed

\subsection{The Influence of the Boundary}
\label{Section_3.4}

While Section \ref{Section_3.3} states the limit of the local objective function, this must be combined with the limit of the localized parameter space in order to derive the limit of the estimator. 
When the parameter space has a boundary, this may influence the limit theory through the limit of the localized parameter space. 
It is well-known that, under strong identification, this influence occurs when the sequence is local to the boundary at the $n^{-1/2}$ rate. 
We show that, under semi-strong identification, this influence occurs when the sequence converges to the boundary at a slower rate, and under weak identification, this influence is unavoidable even if the sequence does not converge to the boundary. 

We need to formalize the definition of the setwise limit of a localized parameter space. 
The type of setwise limit that we use is called Kuratowski convergence. 
We say that a sequence of sets, $\Psi_n$, satisfies $\Psi_n\rightarrow_K \Psi$ for some set, $\Psi$, if for every compact set, $K$, $\max\left(\vec d(\Psi_n\cap K,\Psi),\vec d(\Psi\cap K, \Psi_n)\right)\rightarrow 0$.\footnote{Recall $\vec d(A,B)=\max\left(\sup_{a\in A}\inf_{b\in B}d(a,b),0\right)$. Lemma B.1 in the supplemental materials shows that this definition of Kuratowski convergence is the same as the usual definition given in, for example, Definition 1.1.1 in \cite{AubinFrankowska1990}.} 
Kuratowski convergence is a very weak way to define setwise convergence, and in particular is weaker than Hausdorff convergence.\footnote{A simple example of a sequence of sets that Kuratowski converges but does not Hausdorff converge is $\Psi_n=[-n,n]$ and $\Psi=\R$.} 
Kuratowski convergence can be thought of as a ``pointwise'' version of Hausdorff convergence: the sense in which $\Psi_n$ approximates $\Psi$ does not have to hold uniformly over the ambient space.\footnote{Also note that Kuratowski and Hausdorff convergence are equivalent for subsets of a compact metric space. When the metric space is unbounded, then the definitions of convergence are different. This is our concern because the limit of the localized parameter space is usually unbounded.} 

The following assumption is used under strong and semi-strong identification. 
It specifies the rate at which the true sequence of parameter values must converge to the boundary in order for the boundary to be relevant to the limit theory. 
\begin{assumption}\label{AssumptionTangentCone}\mbox{} For a given sequence $\theta_n=(\pi_n,\beta_n)\rightarrow \theta_\ast=(\pi_\ast,\beta_\ast)$: 
\begin{enumerate}[label=(\alph*)]
\item There exists a closed set, $\Psi\subseteq \R^{d_\theta}$, such that $\{(\sqrt{n}(\pi-\pi_n),\sqrt{n}(\beta-\beta_n)/a_n)|$ $(\pi,\beta)\in\Theta\}\rightarrow_K\Psi$. 
\item $\Psi$ is convex. 
\end{enumerate}
\end{assumption}
\textbf{Remarks:} 
\textbf{3.11.} Under strong identification, Assumption \ref{AssumptionTangentCone}(a) is related to conditions from the parameter-on-the-boundary literature. 
That literature generally requires stronger conditions on the shape of the parameter space, such as Clarke or Chernoff regularity, to ensure the localized parameter space can be approximated by a tangent cone; see \cite{Geyer1994} or \cite{Andrews1999}. 
A new argmax theorem in \cite{CoxArgmax} shows that these stronger conditions are not needed for the purpose of the argmax theorem; only Kuratowski convergence is needed. 
Sufficient conditions for Assumption \ref{AssumptionTangentCone}(a) are very weak. 
A general property of Kuratowski convergence is that every sequence of sets has a subsequence that Kuratowski converges to some limit set; see Theorem 1.1.7 in \cite{AubinFrankowska1990}. 
This means that, in general, all that is required to show Kuratowski convergence is to show that the sequence of sets does not oscillate between two different limit sets. 
This can be guaranteed by choosing the sequence $\{\theta_n\}_{n=1}^{\infty}$ appropriately. 
For this reason, Assumption \ref{AssumptionTangentCone}(a) is not needed for Theorem \ref{Theorem3}, below. 
Also note that Assumption \ref{AssumptionTangentCone}(b) is only needed for the estimation result in Theorem \ref{Theorem1}, and not for the inference results in Section 4. 

\textbf{3.12.} If $\theta_n$ converges to the interior of $\Theta$, then Assumption \ref{AssumptionTangentCone}(a) is satisfied with $\Psi=\R^{d_\theta}$. 
If $\theta_n$ is local to the boundary at the $\sqrt{n}$ rate, then Assumption \ref{AssumptionTangentCone}(a) is satisfied with $\Psi$ being a cone or a translation of a cone. 
Typically, when $\Theta=\{\theta\in\R^{d_\theta}| \ell(\theta)\le 0\}$, Assumption \ref{AssumptionTangentCone} is satisfied with $\Psi=\{\psi\in\R^{d_\theta}| \bar\ell + \partial_\theta \ell(\theta_\ast)\psi\le 0\}$, where $\bar\ell=\lim_{n\rightarrow \infty}\sqrt{n}\ell(\theta_n)$ if the limit exists.\footnote{The limit can be taken componentwise, as some components of $\bar\ell$ may be $-\infty$.} 
This is true in Examples 1 and 2, where $\ell(\theta)$ is defined in (\ref{Ex1_ell_function}) and (\ref{GARCH_ell_definition}), respectively. 

\textbf{3.13.} Under semi-strong identification, $\beta_n$ may converge to the boundary at an arbitrarily slow rate and still satisfy Assumption \ref{AssumptionTangentCone}(a) with $\Psi\neq\R^{d_\theta}$. 
Practically, this means that the influence of the boundary is salient for parameter values further from the boundary than under strong identification. 
Typically, when $\Theta=\{\theta\in\R^{d_\theta}| \ell(\theta)\le 0\}$ and $\Theta$ satisfies Assumption \ref{AssumptionPSS}, Assumption \ref{AssumptionTangentCone} is satisfied with $\Psi=\R^{d_\pi}\times \{\psi_2\in\R^{d_\beta}| \bar\ell + \partial_\beta \ell(\theta_\ast)\psi_2\le 0\}$, where $\bar\ell=\lim_{n\rightarrow \infty}\sqrt{n}\ell(\theta_n)/a_n$ if the limit exists. 
Relative to the limiting set under strong identification, the limiting set under weak identification is rotated as if the inequalities only enforce in the $\beta$ directions. 
This is because the local parameter space is rescaled at different rates for the $\pi$ and $\beta$ parameters. 
\qed \medskip

Assumption \ref{AssumptionTangentCone} is not needed for weak identification. 
The corresponding result for weak identification follows from Assumption \ref{AssumptionPSS}. 
\begin{lemma}\label{Lemma4}
If $\Theta$ satisfies Assumption \ref{AssumptionPSS}, then $\{(\sqrt{n}(\pi-\pi_n),\beta)|(\pi,\beta)\in\Theta\}\rightarrow_K \R^{d_\pi}\times \mathcal{B}_\ast$. 
\end{lemma}

\textbf{Remark:} 
\textbf{3.14.} Lemma \ref{Lemma4} describes the setwise limit of the local parameter space under weak identification. 
As under semi-strong identification, the limit is a product between the $\pi$ and $\beta$ directions because the local parameter space is rescaled at different rates. 
Under weak identification, the bounds influence the limit theory through $\mathcal{B}_\ast=\{\beta\in\R^{d_\beta}| \ell(\pi_\ast,\beta)\le 0\}$. 
The true sequence of parameter values does not need to converge to the boundary for the bounds to be relevant. 
\qed

\subsection{Limit Theory}

For the limit theory under weak identification, one additional assumption is needed. 
\begin{assumption}
\label{AssumptionMIN}
Almost surely, the sample path of $\tilde{q}^W(\beta)$ is uniquely minimized over $\mathcal{B}_\ast$. 
Denote the unique minimizing point by $\beta_{\text{MIN}}$. 
\end{assumption}
\textbf{Remarks:}
\textbf{3.15.} As we know from Lemma \ref{Lemma3}, $\tilde{q}^W(\beta)$ is the limit of the local objective function under weak identification after concentrating out the $\psi_1$ parameters: 
$\tilde{q}^W(\beta)=\inf_{\psi_1\in\R^{d_\pi}}q^W(\psi_1,\beta)$. 
As stated below, $\beta_{\text{MIN}}$ is the limit of $\hat{\beta}_n$ under weak identification. 
Assumption \ref{AssumptionMIN} ensures the limit is well-defined. 
This assumption is only used for the estimation results. 
It is not needed for the inference results in Section 4. 

\textbf{3.16.} Theorem 4 in \cite{Cox2020} provides sufficient conditions for verifying Assumption \ref{AssumptionMIN} based on the derivative of $\tilde{q}^W(\beta)$ with respect to $Y$. 
We use a modification of that theorem to verify Assumption \ref{AssumptionMIN} in the latent factor model examples. 
\qed \medskip

The following theorem states the limit theory for $\hat\theta_n$ under all three identification strengths. 

\begin{theorem}\label{Theorem1}
Suppose Assumptions \ref{AssumptionNCIS}, \ref{AssumptionRFO}, \ref{AssumptionRFI}, \ref{AssumptionPSS}, \ref{AssumptionQE}, and \ref{AssumptionRFL} hold for a given sequence $\{\theta_n\}_{n=1}^\infty$ such that $\theta_n\rightarrow \theta_\ast$.  
\begin{enumerate}[label=(\alph*)]
\item Under Assumptions 5(S) and \ref{AssumptionTangentCone}, 
\begin{align*}
\sqrt{n}(\hat\theta_n-\theta_n)\rightarrow_d &\underset{\psi\in\Psi}{\argmin} q^S(\psi). 
\end{align*}
\item Under Assumptions 5(SS) and \ref{AssumptionTangentCone}, 
\begin{align*}
\left(\begin{array}{c}\sqrt{n}(\hat{\pi}_n-\pi_n)\\\sqrt{n}(\hat{\beta}_n-\beta_n)/a_n\end{array}\right)\rightarrow_d &\underset{\psi\in\Psi}{\argmin} q^S(\psi). 
\end{align*}
\item Under Assumptions 5(W) and \ref{AssumptionMIN},  
\[
\left(\begin{array}{c}\sqrt{n}(\hat{\pi}_n-\pi_n)\\\hat{\beta}_n\end{array}\right)\rightarrow_d\underset{(\psi_1,\beta)\in \R^{d_\pi}\times \mathcal{B}_\ast}{\argmin} q^W(\psi_1,\beta)=\left(\begin{array}{c}Z_1(\beta_{\text{MIN}})\\\beta_{\text{MIN}}\end{array}\right).
\]
\end{enumerate}
\end{theorem}
\textbf{Remarks:} 
\textbf{3.17.} Under strong identification, and if $\Psi=\R^{d_\theta}$, the limit in part (a) simplifies to $Z=-J\inv D'_\ast Y\sim N(0,J\inv D'_\ast V D_\ast J\inv)$, which coincides with the usual asymptotic normal distribution of a minimum distance estimator. 
If $\Psi\neq\R^{d_\theta}$, then the limit in part (a) agrees with limit theory in \cite{Andrews1999}. 

\textbf{3.18.} Under semi-strong identification, part (b) shows that the rate of convergence of $\hat{\beta}_n$ is diminished by $a_n$, the weakness of the sequence. 
In addition, the limiting distribution differs, in general, from the limiting distribution in part (a) because Assumption \ref{AssumptionTangentCone}(a) is satisfied with a different $\Psi$. 
This is a new challenge for uniformly valid inference. 
(Existing weak-identification-robust methods in AC12 and \cite{HanMcCloskey2019} require the limiting distribution under strong and semi-strong identification to be the same.) 
Because the boundary is relevant at the $\sqrt{n}/a_n$ rate, which can be arbitrarily slow, is it hard to consistently estimate the effect. 
We propose a solution in Section \ref{Section4}; see Remark 4.21. 

\textbf{3.19.} Part (c) shows that nonstandard limit theory applies under weak identification. 
In this case, $\hat{\beta}_n$ is inconsistent and converges to the argmin of a stochastic process over the identified set, $\beta_{\text{MIN}}$, and $\hat{\pi}_n$ is $\sqrt{n}$-consistent, but with a nonstandard limiting distribution that can be described as a Gaussian stochastic process function of $\beta_{\text{MIN}}$. 

\textbf{3.20.} The proof of Theorem \ref{Theorem1} verifies the conditions of a recent argmax theorem in \cite{CoxArgmax} that allows the domain (the local parameter space) to change with the sample size. 
The key innovation in the proof is the existence of $\Pi_n\times \mathcal{B}_n$, a sequence of neighborhoods of the identified set, to deal with an informative boundary of the parameter space. 
\qed 

\section{Identification-Robust Inference}
\label{Section4}

This section provides a recommendation for identification-robust hypothesis testing. 
The general strategy is to derive the limit of the quasi-likelihood ratio statistic along sequences of parameters that can be strong, semi-strong, or weak. 
Then, a critical value can be calculated using the quantiles of the limit. 

We consider hypotheses of the form 
\begin{equation}
H_0: r_1(\pi)=0_{d_{r_1}} \text{ and }r_2(\beta)=0_{d_{r_2}}. \label{H0}
\end{equation}
Expression (\ref{H0}) requires each restriction to only depend on $\pi$ or $\beta$ but not both. 
This type of hypothesis is assumed in AC12. 
They suggest that, if it is not satisfied in the original parameterization, it may be satisfied by a reparameterization.\footnote{Finding a reparameterization that satisfies (\ref{H0}) and the product space structure required by Assumption B1 in AC12 can be prohibitively challenging. This paper, by relaxing the product space requirement, makes finding such a reparameterization easier.}
\cite{HanMcCloskey2019} test more general hypotheses by explicitly allowing for the possibility the test statistic may diverge under the null when (\ref{H0}) is not satisfied. We leave an analysis of the limit theory for weak identification with bounds when (\ref{H0}) is not satisfied to future research. \medskip

\noindent\textbf{Examples 1 and 2, Continued.} 
The simulations of Examples 1 and 2 test $H_0: \beta=\beta_0$, for some hypothesized value, $\beta_0$. 
In Example 1, this amounts to testing the variance of the factor. 
Table 6 in \cite{CoxWeakIdFactor} gives reparameterizations for other common hypotheses in factor models with one factor that satisfy (\ref{H0}). 
In Example 2, this amounts to testing the GARCH coefficient. 
To test the ARCH coefficient, one can use $\alpha(\pi,\beta)$ to reparameterize the model in terms of $\alpha$. 
\qed\medskip

Let $r(\pi,\beta)=(r_1(\pi),r_2(\beta))$ and let $d_r=d_{r_1}+d_{r_2}$ denote the number of restrictions. 
Typically, $r_1(\pi)$ and $r_2(\beta)$ restrict a subvector of $\pi$ and/or $\beta$ to take a particular value. 
In that case, the restricted subvectors are null-imposed parameters while the unrestricted parameters are the nuisance parameters. 
Below, we make an important distinction depending on whether $r_2(\beta)$ uniquely determines the value of $\beta$ (so all the nuisance parameters are strongly identified), which we denote by ``W1'' in the following. 
When a component of $\beta$ is not uniquely determined by $r_2(\beta)$, then at least one nuisance parameter is potentially weakly identified, which we denote by ``W2'' in the following. 

\subsection{Quasi-Likelihood Ratio Statistic}

We test $H_0$ using a quasi-likelihood ratio (QLR) statistic. 
Let $\Theta^r=\{\theta\in\Theta| r(\theta)=0\}$ denote the null-imposed parameter space. 
Define the QLR statistic to be 
\[
QLR_n=2n\left(\inf_{\theta\in\Theta^r} Q_n(\theta)-\inf_{\theta\in\Theta} Q_n(\theta)\right). 
\]
We test $H_0$ by comparing $QLR_n$ to a critical value based on the asymptotic distribution of $QLR_n$ under weak and strong identification. 

Let $\theta_n=(\pi_n,\beta_n)\rightarrow \theta_\ast=(\pi_\ast, \beta_\ast)$ and $\xi_n\rightarrow\xi_\ast$ be a sequence of parameters, as in the previous section. 
Let $\mathcal{B}^r(\pi)=\{\beta\in\mathcal{B}(\pi)| r_2(\beta)=0\}$ denote the null-imposed cross section for $\beta$. 
Then, $\mathcal{B}^r_\ast=\mathcal{B}^r(\pi_\ast)$ is the subset of the identified set that satisfies the null hypothesis. 
The limit theory depends on whether $\mathcal{B}^r_\ast$ is a singleton. 
When $\mathcal{B}^r_\ast$ is a singleton, all the nuisance parameters are strongly identified under the null (case W1). 
Otherwise, there exist weakly identified nuisance parameters under the null (case W2). 
The following assumption formalizes this distinction and imposes regularity conditions on $r(\theta)$. 
To aid the reader, the brackets indicate types of sequences associated with each assumption for Theorem \ref{Theorem3}, below. 

\begin{assumption}\label{AssumptionHR}\mbox{}  For a given sequence $\theta_n=(\pi_n,\beta_n)\rightarrow \theta_\ast=(\pi_\ast,\beta_\ast)$: 
\begin{enumerate}[label=(\alph*)]
\item \mbox{}[SS and W2] $r(\theta)$ is continuously differentiable, and $R_1:=\partial_\pi r_1(\pi_\ast)$ has full rank, $d_{r_1}$. 
\item \mbox{}[SS and W1] $r_2(\beta)=0$ implies $\beta=\beta_\ast$. 
\item \mbox{}[SS and W2] There exists an $M>0$ such that for all $\pi_1$ and $\pi_2$ in a neighborhood of $\pi_\ast$ satisfying $r_1(\pi_1)=r_1(\pi_2)=0$, $d_H(\mathcal{B}^r(\pi_1),\mathcal{B}^r(\pi_2))\le M\|\pi_1-\pi_2\|$. 
\item For a closed set, $\Psi^r\subseteq\R^{d_\theta}$, $\{(\sqrt{n}(\pi-\pi_n),\sqrt{n}(\beta-\beta_n)/a_n)| (\pi,\beta)\in\Theta^r\}\rightarrow_K \Psi^r$. 
\end{enumerate}
\end{assumption}
\textbf{Remarks:} 
\textbf{4.1.} Part (b) states that there are no weakly identified nuisance parameters under the null hypothesis. 
Otherwise, when there are weakly identified nuisance parameters under the null hypothesis, we assume part (c), which guarantees the null-imposed cross-sections are Lipschitz in the Hausdorff distance, similar to Assumption \ref{AssumptionPSS}(b). 

\textbf{4.2.} Part (d) is the null-imposed version of Assumption \ref{AssumptionTangentCone}(a). 
As for Assumption \ref{AssumptionTangentCone}(a), it is used under strong and semi-strong identification. 
Part (d) is usually satisfied with $\Psi^r=\{(\psi_1,\psi_2)\in\Psi| R_1\psi_1=0_{d_{r_1}} \text{ and }R_2\psi_2=0_{d_{r_2}}\}$, where $R_1$ is defined in part (a) and $R_2:=\partial_\beta r_2(\beta_\ast)$. 
Under semi-strong identification, $\Psi^r$ inherits the nonstandard features of $\Psi$: the inequalities are rotated as if they only enforce in the $\beta$ directions, and the boundary is relevant at an arbitrarily slow rate. 
\qed\medskip

Assumption \ref{AssumptionHR}(d) gives the limit of the localized parameter space under Assumptions 5(S) or (SS). 
It is also used under Assumption 5(W) for case W1. 
In case W2, the corresponding result follows from Assumption \ref{AssumptionHR}(c), as the following lemma shows. 
Let $R_1^\perp=\{\psi_1\in\R^{d_\pi}| R_1\psi_1=0_{d_{r_1}}\}$. 

\begin{lemma}\label{Lemma5}
If $\Theta^r$ satisfies Assumption \ref{AssumptionHR}(a,c), then $\{(\sqrt{n}(\pi-\pi_n),\beta)|(\pi,\beta)\in\Theta^r\}\rightarrow_K R^\perp_1\times\mathcal{B}^r_\ast$. 
\end{lemma}

The following definition is used to state the limit of $QLR_n$ under W2. 
For every $\beta\in\mathcal{B}_1$, let 
\begin{align}
\tilde q^{W,r}(\beta)&=\tilde q^W(\beta)+Z_1(\beta)'R'_1\left(R_1J_{11}(\beta)\inv R'_1\right)^{-1}R_1Z_1(\beta)/2. \label{4.2}
\end{align}
This formula characterizes the contribution of the $r_1(\pi)$ restrictions to the concentrated limit objective function from Remark 3.15. 
The following theorem characterizes the limiting distribution of $QLR_n$. 

\begin{theorem}\mbox{} \label{Theorem2}
Suppose Assumptions \ref{AssumptionNCIS}, \ref{AssumptionRFO}, \ref{AssumptionRFI}, \ref{AssumptionPSS}, \ref{AssumptionQE}, and \ref{AssumptionRFL} hold for a given sequence $\{\theta_n\}_{n=1}^\infty$ such that $\theta_n\rightarrow \theta_\ast$, and suppose $r(\theta_n)=0$ for all $n$. 
\begin{enumerate}[label=(\alph*)]
\item Under Assumptions 5(S), \ref{AssumptionTangentCone}(a), and \ref{AssumptionHR}(d), 
\[
QLR_n\rightarrow_d QLR_\ast^S := \inf_{\psi\in\Psi^r}(Z-\psi)'J(Z-\psi)-\inf_{\psi\in\Psi}(Z-\psi)'J(Z-\psi). 
\]
\item Under Assumptions 5(SS), \ref{AssumptionTangentCone}(a), and \ref{AssumptionHR}(a,(b or c),d), 
\[
QLR_n\rightarrow_d QLR_\ast^{S} =\inf_{\psi\in\Psi^r}(Z-\psi)'J(Z-\psi)-\inf_{\psi\in\Psi}(Z-\psi)'J(Z-\psi). 
\]
\item Under Assumptions 5(W) and \ref{AssumptionHR}(b,d), 
\begin{align*}
QLR_n\rightarrow_d QLR^{W1}_\ast:=&\inf_{(\psi_1,\psi_2)\in\Psi^r}(Z_1(\beta_\ast)-\psi_1)'J_{11}(\beta_\ast)(Z_1(\beta_\ast)-\psi_1)\\
&-Z_1(\beta_\ast)'J_{11}(\beta_\ast)Z_1(\beta_\ast)-\inf_{\beta\in\mathcal{B}_\ast}2\tilde q^W(\beta). 
\end{align*}
\item Under Assumptions 5(W) and \ref{AssumptionHR}(a,c), 
\[
QLR_n\rightarrow_d QLR^{W2}_\ast:=\inf_{\beta\in\mathcal{B}^r_\ast}2\tilde q^{W,r}(\beta)-\inf_{\beta\in\mathcal{B}_\ast}2\tilde q^W(\beta). 
\]
\end{enumerate}
\end{theorem}
\textbf{Remarks:} 
\textbf{4.3.} The limit theory for part (a) agrees with the limit theory derived in \cite{Andrews2001} for the QLR statistic (without a parameter that is not identified under the null). 
The limit reduces to $\chi^2_{d_r}$ when $\Psi=\R^{d_\theta}$, $\Psi^r$ is a subspace of dimension $d_\theta-d_r$, and $H=V$. 
The limit theory for part (b) looks the same as part (a) but is different because Assumptions \ref{AssumptionTangentCone}(a) and \ref{AssumptionHR}(d) are satisfied with different $\Psi$ and $\Psi^r$. 

\textbf{4.4.} Parts (c) and (d) state limit theory for the QLR statistic under weak identification. 
For both, the limit theory of the unrestricted part is the same. 
Under Assumption \ref{AssumptionHR}(b,d), the limit theory for the null-imposed part resembles the strong identification case, while under Assumption \ref{AssumptionHR}(a,c), the limit theory for the null-imposed part resembles the weak identification case. 
\qed

\subsection{Inference Recommendation}

We recommend testing $H_0$ with the following steps: (1) Calculate $QLR_n$. (2) Calculate a WIR critical value. 
(3) Reject if $QLR_n$ exceeds the critical value. 
The WIR critical value is based on the quantiles of the distribution of $QLR^L_\ast$, where $L\in\{S,W1,W2\}$. 
This critical value uses a quantile of $QLR^S_\ast$ when identification is strong and a quantile of $QLR^{W1}_\ast$ or $QLR^{W2}_\ast$ when identification is weak. 
At the end of this section, we give instructions for simulating the quantiles of $QLR^L_\ast$. 
Notice that the distribution of $QLR^L_\ast$ depends on the objects: $\theta_\ast$, $H$, $V$, $c(\cdot)$, $\Psi$, and $\Psi^r$. 
Recall $\theta_\ast$ is the limit of the sequence of parameter values, $\theta_n$, $H$ and $V$ are defined in Assumptions \ref{AssumptionQE} and \ref{AssumptionRFL}, $c(\cdot)$ is defined in Assumption 5(W), $\Psi$ is defined in Assumption \ref{AssumptionTangentCone}(a), and $\Psi^r$ is defined in Assumption \ref{AssumptionHR}(d). 
The distribution also depends on $\mathcal{B}_\ast$ and $\mathcal{B}^r_\ast$, but we write these as $\mathcal{B}(\pi_\ast)$ and $\mathcal{B}^r(\pi_\ast)$, respectively, making the dependence on $\pi_\ast$ explicit. 

Let $F_L(\cdot; \theta_\ast, H, V, c(\cdot), \Psi, \Psi^r)$ denote the distribution function of $QLR^L_\ast$, for $L\in\{S, W1, W2\}$.\footnote{Note that $F_L(\cdot)$ does not depend on all of the objects for particular values of $L$. Specifically, $F_S(\cdot)$ does not depend on $c(\cdot)$, $F_{W1}(\cdot)$ does not depend on $\Psi$, and $F_{W2}(\cdot)$ does not depend on $\Psi$ or $\Psi^r$. We abuse notation by dropping the irrelevant arguments when a particular value of $L$ is considered.}
Let $\alpha\in(0,1)$ denote the nominal size of the test, and let $q_L(1-\alpha; \theta_\ast, H, V, c(\cdot),$ $\Psi,$ $\Psi^r)$ denote the $1-\alpha$ quantile of the $F_L(\cdot;\theta_\ast, H, V, c(\cdot),$ $\Psi, \Psi^r)$ distribution.\footnote{The $1-\alpha$ quantile is defined as $q(1-\alpha):=\inf\{u\in\R|F(u)\ge 1-\alpha\}$.} 
We first state a technical assumption that requires the quantile to be a continuity point of the distribution. 
\begin{namedassumption}[11($\bm{\alpha}$,$\bm{L}$)]
For a given $\theta_\ast$, $H$, $V$, $c(\cdot)$, $\Psi$, and $\Psi^r$, $q_L(1-\alpha;$ $\theta_\ast,$ $H, V,$ $c(\cdot),$ $\Psi, \Psi^r)$ is a continuity point of $F_L(\cdot; \theta_\ast, H, V, c(\cdot),$ $\Psi, \Psi^r)$. 
\end{namedassumption}
\stepcounter{assumption}
\textbf{Remark:} 
\textbf{4.5.} Assumption 11($\alpha$,$L$) ensures the probability the test statistic is equal to the critical value is zero in the limit. 
We use Assumption 11($\alpha$,$L$) for different values of $\alpha$ and $L\in\{S,W1,W2\}$ in Theorem \ref{Theorem3} below. 
\qed\medskip 

In order to estimate the quantiles of the distribution of $QLR^L_\ast$, we need estimators of the parameters indexing $q_L(1-\alpha; \theta_\ast, H, V, c(\cdot),$ $\Psi,$ $\Psi^r)$ for $L\in\{S,W1,W2\}$. 
In what follows, we describe estimators for these parameters and state high-level conditions for uniform validity of the resulting critical value. 
The high-level conditions are verified in factor models with one or two factors in the supplemental materials. 

We start with parameters that can be estimated from the reduced-form model: $\pi$, $H$, and $V$. 
We also consider $\beta$, a structural parameter that usually cannot be consistently estimated under weak identification but can be consistently estimated under strong and semi-strong identification. 
The following assumption places conditions on estimators of these parameters. 

\begin{assumption}\mbox{}\label{AssumptionFSab} There exist estimators $\hat\pi$, $\hat\beta$, $\hat H$, and $\hat V$ that satisfy the following conditions under the given sequence, $\theta_n\rightarrow\theta_\ast$ and $\xi_n\rightarrow\xi_\ast$. 
\begin{enumerate}[label=(\alph*)]
\item \mbox{}[All sequences] $\sqrt{n}(\hat\pi-\pi_n)=O_p(1)$, $r_1(\hat\pi)=0$, $\hat H\rightarrow_p H$, and $\hat V\rightarrow_p V$. 
$\hat H$ and $\hat V$ are positive definite and symmetric almost surely. 
\item \mbox{}[S and SS] $\hat\beta\rightarrow_p \beta_\ast$. 
\end{enumerate}
\end{assumption}
\textbf{Remarks:}
\textbf{4.6.} Part (a) ensures consistency of $\hat H$, $\hat V$, and $\hat\pi$. 
It is assumed for all identification strengths, strong, semi-strong, and weak. 
Part (a) also assumes $\hat\pi$ satisfies the null in the sense that $r_1(\hat\pi)=0$. 
Part (b) is used for consistent estimation of $\beta$. 
It is only assumed for strong and semi-strong identification. 

\textbf{4.7.} 
There are a few options for estimators $\hat\pi$ and $\hat\beta$ that satisfy Assumption \ref{AssumptionFSab}. 
A convenient one is to let $\breve\Theta^r=\{\theta\in\breve\Theta| r(\theta)=0_{d_r}\}$ and define $\breve\theta_n=(\breve\pi_n,\breve\beta_n)$ using (\ref{MEstimator}) with $\Theta$ replaced by $\breve\Theta^r$. 
Note that $\breve\Theta^r$ imposes the null hypothesis but does not impose the bounds. 
This is convenient because $\breve\theta_n$ is asymptotically normal under strong and semi-strong identification, as well as under weak identification if Assumption \ref{AssumptionHR}(b) holds. 
We use $\breve\theta_n$ in the simulations. 
\qed\medskip

We next consider estimating $c(\cdot)$, defined in Assumption 5(W). 
A consistent estimator usually does not exist under weak identification. 
Instead, we follow a general Bonferroni approach developed in \cite{McCloskey2017} that uses a first-stage confidence set for $c(\cdot)$. 
Fix a first-stage level $\alpha_{c}\in[0,\alpha)$ and let $\hat \Pi$ denote a set of values of $\pi$. 
For every $\pi_\ast\in\hat\Pi$, let 
\begin{equation}
\hat c(\beta;\pi_\ast,\beta_\ast)=\sqrt{n}(\tau(\pi_\ast,\beta)-\tau(\pi_\ast,\beta_\ast)).  \label{dform}
\end{equation}
For a fixed value of $\pi_\ast$ and $\beta_\ast$, $\hat c(\cdot;\pi_\ast,\beta_\ast)$ approximates $c(\cdot)$. 

\begin{assumption}\mbox{}\label{AssumptionFScd} There exists a set $\hat\Pi$ such that $\hat\pi$ and $\hat\Pi$ satisfy the following conditions under the given sequence, $\theta_n\rightarrow\theta_\ast$ and $\xi_n\rightarrow\xi_\ast$. 
\begin{enumerate}[label=(\alph*)]
\item \mbox{}[W] For every $\epsilon>0$, 
\[
\hspace{-0.35cm}\liminf_{n\rightarrow\infty}P_{\theta_n,\xi_n}\hspace{-0.6mm}\left(\inf_{\hat\pi_\ast\in\hat \Pi}\inf_{\hat\beta_\ast\in\mathcal{B}^r(\hat\pi)}\left(\hspace{-0.4mm}\|\hat\beta_\ast\hspace{-0.4mm}-\hspace{-0.4mm}\beta_\ast\|+\sup_{\beta\in \mathcal{B}_1}\|\hat c(\beta;\hat\pi_\ast,\hat\beta_\ast)\hspace{-0.4mm}-\hspace{-0.4mm}c(\beta)\|\right)\le\epsilon\right)\hspace{-0.4mm}\ge\hspace{-0.4mm} 1-\alpha_c,
\]
where $c(\beta)$ is defined in Assumption 5(W), and $\mathcal{B}_1$ is defined in Lemma \ref{LemmaCON}. 

\item \mbox{}[SS] There exists a neighborhood, $U$, of $\beta_\ast$ such that 
\begin{align*}
\inf_{\hat \pi_\ast\in\hat \Pi}\inf_{\hat\beta_\ast\in\mathcal{B}^r(\hat\pi)}\left(\frac{\sqrt{n}\|\hat\beta_\ast-\beta_n\|}{a_n}\right.&+\sup_{\beta\in \mathcal{B}_1}\|\frac{a_n}{\sqrt{n}}\hat c(\beta;\hat\pi_\ast,\hat\beta_\ast)-c(\beta)\|\\
&\left.+\sup_{\beta\in U}\|\frac{a_n}{\sqrt{n}}\partial_\beta \hat c(\beta;\hat\pi_\ast,\hat\beta_\ast)-C(\beta)\|\right)\rightarrow_p 0, 
\end{align*}
where $c(\beta)$, $C(\beta)$, and $a_n$ are defined in Assumption 5(SS) and $\mathcal{B}_1$ is defined in Lemma \ref{LemmaCON}. 
\end{enumerate}
\end{assumption}
\textbf{Remarks:}
\textbf{4.8.} Part (a) is only used for weak identification. 
It ensures the set of all $\hat c(\beta; \hat\pi_\ast,\hat\beta_\ast)$ for $\hat\pi_\ast\in\hat\Pi$ and $\hat\beta_\ast\in \mathcal{B}^r(\hat\pi)$ 
contains an approximation to the true $c(\beta)$ with probability at least $1-\alpha_c$ asymptotically. 
(It also requires the $\hat\beta_\ast$ that is used in the approximation of $c(\beta)$ to be consistent for $\beta_\ast$.) 
Part (b) is only used for semi-strong identification. 
It says that $\hat c(\beta;\hat\pi_\ast,\hat\beta_\ast)$ for $\hat\pi_\ast\in\hat\Pi$ and $\hat\beta_\ast\in\mathcal{B}^r(\hat\pi)$ is capable of approximating both $c(\beta)$ and $C(\beta)$ (by its derivative) simultaneously with probability approaching 1. 
(It also requires the $\hat\beta_\ast$ that is used in the approximation of $c(\beta)$ and $C(\beta)$ to be consistent for $\beta_\ast$ at the $\sqrt{n}/a_n$ rate.) 

\textbf{4.9.} Verifying Assumption \ref{AssumptionFScd} comes down to $\hat\Pi$ and verifying that it contains values sufficiently close to $\pi_n$. 
To get intuition, suppose $\tau(\pi,\beta)=\pi_1\tilde\tau(\pi,\beta)$, so Assumption A in AC12 is satisfied; see Remark 2.2. 
In this case, $\hat\Pi$ can be constructed using a $1-\alpha_c$ confidence set for $\pi_1$, together with consistent estimators of the other elements of $\pi$. 
If $\hat\pi$ is asymptotically normal under weak identification, then the $1-\alpha_c$ confidence set for $\pi_1$ can be constructed in the usual way with standard normal quantiles and a consistent estimator of the standard error of $\hat\pi$.\footnote{\cite{HanMcCloskey2019} construct a similar confidence set for the local parameter under weak identification. They use the original estimator, $\hat\pi_n$, which has the problem that it is not asymptotically normal under weak identification (and its distribution depends on the local parameter). This problem can be solved in minimum distance models because there usually exists an asymptotically normal estimator for $\pi$. If Assumption \ref{AssumptionHR}(b) is satisfied, then $\breve\pi_n$ from Remark 4.7 is asymptotically normal. Otherwise, the reduced-form estimator for $\delta=(\pi,\tau)$ that minimizes $T_n(\delta)$ over an open expansion of $\delta(\Theta)$ is asymptotically normal.} 
Below we give formulas for $\hat\Pi$ in Examples 1 and 2 that satisfy Assumption \ref{AssumptionFScd}. 

\textbf{4.10.} The role of $\alpha_c$ is to capture the finite-sample probability that $\hat\Pi$ does not contain values sufficiently close to $\pi_n$. 
It is possible to satisfy Assumption \ref{AssumptionFScd} with $\alpha_c=0$ by increasing the critical value used to construct the confidence set in $\hat\Pi$ as the sample size increases. 
This leads to less conservative inference asymptotically but may not accurately represent the uncertainty from approximating $\pi_n$ in finite sample. 
We cover this possibility in the theoretical results of this paper by not requiring $\alpha_c$ to be positive. 
In the examples and application, we take $\alpha_c=\alpha/10$, a reasonable option suggested in \cite{McCloskey2017}. 
\qed\medskip

We next approximate $\Psi$ and $\Psi^r$ with random sets $\hat\Psi$ and $\hat\Psi^r$. 
Recall $\Psi$ and $\Psi^r$ depend on $\bar\ell=\lim_{n\rightarrow\infty}\sqrt{n}\ell(\theta_n)$, the slackness of the inequalities that define the boundary of $\Theta$; see Remarks 3.12 and 4.2. 
Since we cannot consistently estimate the slackness of an inequality locally at the $\sqrt{n}$ rate, we take the approach of a first-stage confidence set to conservatively approximate $\Psi$ and $\Psi^r$. 
Let $\alpha_\Psi\in[0,\alpha)$ denote the level of a first-stage confidence set. 
The following assumption formalizes the sense in which $\hat\Psi$ and $\hat\Psi^r$ must conservatively approximate $\Psi$ and $\Psi^r$. 
Recall $\vec d(A,B)=\max\left(\sup_{a\in A}\inf_{b\in B}d(a,b),0\right)$ is a directed distance between sets $A$ and $B$. 

\begin{assumption}\mbox{}\label{AssumptionFSef} There exist sets $\hat\Psi$ and $\hat\Psi^r$ that satisfy the following conditions under a given sequence, $\theta_n\rightarrow\theta_\ast$ and $\xi_n\rightarrow\xi_\ast$. 
\begin{enumerate}[label=(\alph*)]
\item \mbox{}[S and SS] For every $K$, compact, and for every $\epsilon>0$, 
\begin{align*}
\liminf_{n\rightarrow\infty}P_{\theta_n,\xi_n}&\left(\vec d\left(\sqrt{n}(\Theta-\theta_n)\cap K,\hat\Psi\right)\le\epsilon\right)\ge 1-\alpha_\Psi.
\end{align*} 

\item \mbox{}[S, SS, and W1] For every $K$, compact, and for every $\epsilon>0$,
\[
\liminf_{n\rightarrow\infty}P_{\theta_n,\xi_n}\left(\vec d\left(\hat\Psi^r\cap K,\sqrt{n}(\Theta^r-\theta_n)\right)\le\epsilon \text{ and }0\in\hat\Psi^r\right)\ge 1-\alpha_\Psi.
\]
\end{enumerate}
\end{assumption}
\textbf{Remarks:} 
\textbf{4.11.} Part (a) requires $\hat\Psi$ to contain points that are arbitrarily close to any point of $\sqrt{n}(\Theta-\theta_n)$ with probability at least $1-\alpha_\Psi$ asymptotically. 
Part (b) requires $\hat\Psi^r$ to only contain points that are arbitrarily close to some point of $\sqrt{n}(\Theta^r-\theta_n)$ with probability at least $1-\alpha_\Psi$ asymptotically. 
Part (a) is used for strong and semi-strong identification, while part (b) is used for strong, semi-strong, and W1 identification. 

\textbf{4.12.} Assumption \ref{AssumptionFSef}(a) is stated with $\sqrt{n}(\Theta-\theta_n)$ instead of $\{(\sqrt{n}(\pi-\pi_n),\sqrt{n}(\beta-\beta_n)/a_n)|(\pi,\beta)\in\Theta\}$ from Assumption \ref{AssumptionTangentCone}. 
This is useful because it is easier to verify. 
Note that $\hat\Psi$ only needs to approximate the local parameter space rescaled at the strongly identified rate. 
It does not need to adapt to approximate the local parameter space rescaled at the semi-strongly identified rate. 
This remark also applies to $\sqrt{n}(\Theta^r-\theta_n)$ in part (b). 

\textbf{4.13.} Sets satisfying Assumption \ref{AssumptionFSef} can be defined using one-sided confidence sets for $\bar\ell$, the slackness parameters on a given collection of inequalities. 
This problem has been considered in the moment inequality literature. 
Below, we define $\hat\Psi$ and $\hat\Psi^r$ in Examples 1 and 2 that satisfy Assumption \ref{AssumptionFSef} by modifying the first-stage confidence for the slackness parameters used in \cite{MoonSchorfheide2009}. 
Assumption \ref{AssumptionFSef} is stated in a general way to cover other approaches for dealing with the slackness parameters, including \cite{AndrewsSoares2010} and \cite{RomanoShaikhWolf2014}. 

\textbf{4.14.} In Assumption \ref{AssumptionFSef}, $\alpha_\Psi$ captures the finite-sample probability that $\hat\Psi$ or $\hat\Psi^r$ do not approximate $\Psi$ or $\Psi^r$. 
As for $\alpha_c$, it is possible to satisfy Assumption \ref{AssumptionFSef} with $\alpha_\Psi=0$ by increasing the critical value used to construct $\hat\Psi$ and $\hat\Psi^r$. 
This leads to less conservative inference asymptotically but may not accurately represent the uncertainty from the approximation of $\Psi$ and $\Psi^r$ in finite sample. 
In the examples and application, we take $\alpha_\Psi=\alpha/10$, following the recommendation in \cite{RomanoShaikhWolf2014}. 
\qed \medskip 

We follow AC12 and use an identification-category-selection (ICS) statistic that distinguishes weak identification from strong identification. 
Let $\hat\kappa$ be a statistic taking values in $[0,1]$. 
\begin{assumption}\label{AssumptionICS} Under the given sequence, $\theta_n\rightarrow\theta_\ast$ and $\xi_n\rightarrow\xi_\ast$, 
\begin{enumerate}[label=(\alph*)]
\item \mbox{}[W] $\hat\kappa\rightarrow_p 1$, or 
\item \mbox{}[S] $\hat\kappa\rightarrow_p 0$.  
\end{enumerate}
\end{assumption}
\textbf{Remark:} 
\textbf{4.15.} Under weak identification, we assume $\hat\kappa$ satisfies Assumption \ref{AssumptionICS}(a), while under strong identification, we assume $\hat\kappa$ satisfies Assumption \ref{AssumptionICS}(b). 
In this way, $\hat\kappa$ is an indicator for weak identification. 
\qed\medskip

Finally, we define the WIR critical value. 
Let $\alpha_{W1}=\alpha-\alpha_c-\alpha_\Psi$, $\alpha_{W2}=\alpha-\alpha_c$, and $\alpha_S=\alpha-2\alpha_{\Psi}$. 
For $L\in\{W1,W2\}$, we assume $0<\alpha_L\le\alpha_S$. 
Let 
\begin{align*}
\widehat{CV}_L&=\hat\kappa \sup_{\hat \pi_\ast\in\hat \Pi}\sup_{\hat\beta_\ast\in \mathcal{B}^r(\hat\pi)}q_L(1-\alpha_L;\hat\pi, \hat\beta_\ast, \hat H, \hat V, \hat c(\cdot;\hat\pi_\ast,\hat\beta_\ast),\hat\Psi^r)\\
&\hphantom{=}+(1-\hat\kappa)q_S(1-\alpha_S; \hat\pi, \hat\beta, \hat H, \hat V, \hat\Psi, \hat\Psi^r).
\end{align*}
This critical value uses $q_S$ when identification is strong and $q_{W1}$ or $q_{W2}$ when identification is weak. 
When identification is semi-strong, all the categories, $q_S$, $q_{W1}$, and $q_{W2}$, automatically adapt to approximate the semi-strong limit of $QLR_n$. 

\begin{namedexample}[1, Continued]
We recommend the following objects to satisfy Assumptions \ref{AssumptionFSab} - \ref{AssumptionICS}. 
We take $\hat\pi$ and $\hat\beta$ to be subvectors of $\breve{\theta}_n=(\breve\rho_{2},\breve\rho_{3},\breve\omega_{1},\breve\omega_{2},\breve\omega_{3},\beta_0)$. 
We weight the objective with $\widehat W_n=\hat{\bar{V}}\inv$, given by the empirical fourth moments of $X_i$, and take $\hat H$ and $\hat V$ to be $\hat{\bar{V}}\inv$. 

For $\hat\Pi$, we use the asymptotic normality of $\breve\rho_3$ to construct a confidence set for $\rho_3$ with coverage probability at least $1-\alpha_c$. 
One can show that $\sqrt{n}(\breve\rho_3-\rho_{3n})\rightarrow_d N(0,e'_2J_{11}(\beta_0)\inv e_2)$, where $\rho_{3n}$ is the second element of $\pi_n$. 
Let 
\[
\hat\Pi=\Big\{(\breve\rho_{2},\rho_3,\breve\omega_{1},\breve\omega_{2},\breve\omega_{3})\Big| \sqrt{n}|\rho_3-\breve\rho_{3}|\le z_{1-\alpha_c/2}\sqrt{e'_2\breve{J}_{11}\inv e_2}\Big\}, 
\]
where $\breve J_{11} = D_1(\breve{\theta}_n)'\hat{\bar V}\inv D_1(\breve{\theta}_n)$. 
This definition of $\hat\Pi$ satisfies Assumption \ref{AssumptionFScd}(a) under weak identification and Assumption \ref{AssumptionFScd}(b) under semi-strong identification. 
Note that a $1-\alpha_c$ confidence interval is used for $\rho_3$ (since $\rho_3$ is the key parameter that determines identification), while only consistent estimators are needed for the other parameters. 

For $\hat\Psi$ and $\hat\Psi^r$, we linearize $\ell(\theta)$ around $\theta_\ast$. 
Let 
\begin{align}
\bar\ell^\pm=&\left[\begin{array}{c}\sqrt{n}\ell_1(\breve{\theta}_n)\pm\left(\partial_\pi \ell_1(\breve{\theta}_n)'\breve{J}_{11}\inv \partial_\pi \ell_1(\breve{\theta}_n)\right)^{1/2}z_{1-\alpha_\Lambda}\\\sqrt{n}\ell_2(\breve{\theta}_n)\pm\left(\partial_\pi \ell_2(\breve{\theta}_n)'\breve{J}_{11}\inv \partial_\pi \ell_2(\breve{\theta}_n)\right)^{1/2}z_{1-\alpha_\Lambda}\end{array}\right] \label{barelldef}
\end{align}
be the endpoints of one-sided confidence sets for $\bar\ell$. 
We use the upper bounds for $\hat\Psi$ and the lower bounds for $\hat\Psi^r$: $\hat\Psi=\{\psi\in\R^6| \min(\bar\ell^-,0)+\partial_\theta\ell(\breve{\theta}_n)\psi\le 0\}$ and $\hat\Psi^r=\{(\psi_1,0)\in\R^6| \min(\bar\ell^+,0)+\partial_\pi\ell(\breve{\theta}_n)\psi_1\le 0\}$. 

Also, $\hat\kappa=\mathds{1}\{n\breve\rho^2_{3}\le \log(n)e'_2\breve{J}_{11}\inv e_2\}$ is an ICS statistic that satisfies Assumption \ref{AssumptionICS}(a) under weak identification and Assumption \ref{AssumptionICS}(b) under strong identification. 
\qed
\end{namedexample}

\begin{namedexample}[2, Continued]
Example 2 uses similar objects as Example 1 
to satisfy Assumptions \ref{AssumptionFSab} - \ref{AssumptionICS}. 
We take $\hat\pi$ and $\hat\beta$ to be the components of $\breve\theta_n=(\breve\pi_n, \beta_0)$, respectively. 
Also, we take $\widehat{W}_n=\hat{\bar{V}}^{-1}$, so we can take $\hat H=\hat V=\widehat{W}_n$. 

For $\hat\Pi$, we use the asymptotic normality of $\breve\pi_n$ to construct a confidence set for $\pi$ with coverage probability at least $1-\alpha_c$. 
Using $\sqrt{n}(\breve\pi_n-\pi_n)\rightarrow_d N(0,J_{11}(\beta_0)^{-1})$, 
let
\[
\hat\Pi=\Big\{\pi\Big| \sqrt{n}|\pi-\breve\pi_n|\le z_{1-\alpha_c/2}\breve{J}_{11}^{-1/2}\Big\}, 
\]
where $\breve{J}_{11}=D_1(\breve{\theta}_n)'\hat{\bar V}\inv D_1(\breve{\theta}_n)$. 
This definition of $\hat\Pi$ satisfies Assumption \ref{AssumptionFScd}(a) under weak identification and Assumption \ref{AssumptionFScd}(b) under semi-strong identification. 

For $\hat\Psi$ and $\hat\Psi^r$, we linearize $\ell(\theta)$ around $\theta_\ast$ just like (\ref{barelldef}) in Example 1. 
We use the upper bounds for $\hat\Psi$ and the lower bounds for $\hat\Psi^r$: $\hat\Psi=\{\psi\in\R^2| \min(\bar\ell^-,0)+\partial_\theta\ell(\breve{\theta}_n)\psi\le 0\}$ and $\hat\Psi^r=\{(\psi_1,0)\in\R^2| \min(\bar\ell^+,0)+\partial_\pi\ell(\breve{\theta}_n)\psi_1\le 0\}$. 

Also, $\hat\kappa=\mathds{1}\{n\breve\pi^2_n\le \log(n)\breve{J}_{11}\inv\}$ is an ICS statistic that satisfies Assumption \ref{AssumptionICS}(a) under weak identification and Assumption \ref{AssumptionICS}(b) under strong identification. 
\qed
\end{namedexample}

We next state the main justification for using $\widehat{CV}_{L}$, which is that it is asymptotically uniformly valid. 
Since the model includes weak identification and bounds on the parameters, this implies the RQLR test robust to weak identification with bounds. 
Up to this point, all results and assumptions have been stated for a fixed sequence, $\{\theta_n,\xi_n\}_{n=1}^{\infty}$. 
Now, we consider classes of sequences of parameters. 
We state the final assumption, a subsequencing condition that guarantees that the three types of sequences, strong, semi-strong, and weak, are sufficient for uniform size control. 
Let $\Gamma$ denote the joint parameter space for $\theta$ and $\xi$. 
Let $\Theta^\dagger\subseteq\Theta$ be compact, and let $\Gamma^\dagger=\{(\theta,\xi)\in\Gamma| \theta\in\Theta^\dagger, \xi\in\Xi^\dagger(\theta), r(\theta)=0\}$, where $\Xi^\dagger(\theta)$ is a set of possible values for $\xi$ for every $\theta\in\Theta^\dagger$. 
We seek to control size uniformly over $\Gamma^\dagger$. 
Let $\Upsilon_L$ for $L\in\{S, SS, W\}$ denote sets of converging sequences, $\{\gamma_n\}_{n=1}^{\infty}\subseteq \Gamma$, such that the following assumption holds. 
\begin{assumption}\label{Assumption14}
For every sequence $\{\gamma^\dagger_n\}_{n=1}^{\infty}\subseteq \Gamma^\dagger$, and for every subsequence, $n_m$, there exists a sequence, $\{\gamma_n\}_{n=1}^{\infty}\in\Upsilon_S\cup \Upsilon_{SS}\cup \Upsilon_W$, and a further subsequence, $n_q$, such that for all $q$, $\gamma_{n_q}=\gamma^\dagger_{n_q}$. 
\end{assumption}
\textbf{Remark:} 
\textbf{4.16.} Assumption \ref{Assumption14} guarantees that sequences contained in $\Upsilon_S\cup \Upsilon_{SS}\cup \Upsilon_W$ are sufficiently ``dense'' in the set of all sequences that they characterize uniformity over $\Gamma^\dagger$. (``Dense''-ness here could be formalized by defining one sequence to belong to the neighborhood of another sequence if they are eventually equivalent along a subsequence.) 
Sufficient conditions for Assumption \ref{Assumption14} are based on defining $\Upsilon_L$ to ensure they include enough sequences of $\{\gamma_n\}_{n=1}^\infty\subseteq\Gamma$. 
Relative to \cite{AndrewsChengGuggenberger2020}, this approach to uniformity separates out the subsequencing condition from the condition that controls rejection probabilities. 
\qed \medskip

The following theorem states that the RQLR test is uniformly valid. 
Note that there are two cases: $L=W1$ indicates that $\beta$ is uniquely determined by $H_0$, while $L=W2$ indicates that $\beta$ is not uniquely determined by $H_0$. 
Which case depends on the type of hypothesis being tested and is known to the researcher. 

\begin{theorem}\mbox{} \label{Theorem3}
Suppose Assumption \ref{Assumption14} holds. Let $L\in\{W1,W2\}$. 
\begin{itemize}
\item[\textendash] Suppose for every $\{\gamma_n\}_{n=1}^{\infty}\in\Upsilon_S$, Assumptions \ref{AssumptionNCIS}, \ref{AssumptionRFO}, \ref{AssumptionRFI}, \ref{AssumptionPSS}, 5(S), \ref{AssumptionQE}, \ref{AssumptionRFL}, 11($\alpha_S$,S), \ref{AssumptionFSab}, \ref{AssumptionFSef}, and \ref{AssumptionICS}(b) hold. 
\item[\textendash] Suppose for every $\{\gamma_n\}_{n=1}^{\infty}\in\Upsilon_{SS}$, Assumptions \ref{AssumptionNCIS}, \ref{AssumptionRFO}, \ref{AssumptionRFI}, \ref{AssumptionPSS}, 5(SS), \ref{AssumptionQE}, \ref{AssumptionRFL}, \ref{AssumptionHR}(a), 11($\alpha_S$,S), \ref{AssumptionFSab}, \ref{AssumptionFScd}(b), and \ref{AssumptionFSef} hold. 
In addition, when $L=W1$, suppose Assumption \ref{AssumptionHR}(b) holds, and when $L=W2$, suppose Assumption \ref{AssumptionHR}(c) holds. 
\item[\textendash] Suppose for every $\{\gamma_n\}_{n=1}^{\infty}\in\Upsilon_W$, Assumptions \ref{AssumptionNCIS}, \ref{AssumptionRFO}, \ref{AssumptionRFI}, \ref{AssumptionPSS}, 5(W), \ref{AssumptionQE}, \ref{AssumptionRFL}, 11($\alpha_L$,$L$), \ref{AssumptionFSab}(a), \ref{AssumptionFScd}(a), and \ref{AssumptionICS}(a) hold. 
In addition, when $L=W1$, suppose Assumptions \ref{AssumptionHR}(b) and \ref{AssumptionFSef}(b) hold, and when $L=W2$, suppose Assumption \ref{AssumptionHR}(a,c) holds.  
\end{itemize}
Then, for $L\in\{W1,W2\}$, 
\[
\limsup_{n\rightarrow\infty} \sup_{\gamma^\dagger\in\Gamma^\dagger} P_{\gamma^\dagger}\left(QLR_n>\widehat{CV}_L\right)\le \alpha. 
\]
\end{theorem}

\textbf{Remarks:} 
\textbf{4.17.} Theorem \ref{Theorem3} shows that the robust critical values control size uniformly over $\gamma\in\Gamma^\dagger$. 
This shows that the RQLR test is identification-robust without the parameter space being a product space. 

\textbf{4.18.} The statement of Theorem \ref{Theorem3} is a bit unusual because the assumptions depend on the type of sequence considered. 
Thus, the statement of Theorem \ref{Theorem3} states the assumptions needed for each type of sequence. 
Not all assumptions are needed for every sequence. 
Only the relevant assumptions for each type of sequence need to be verified in a given model. 
Also notice that Assumptions \ref{AssumptionTangentCone}, \ref{AssumptionMIN}, and \ref{AssumptionHR}(d) are not needed at all. 
This is quite surprising. 

\textbf{4.19.} Under some conditions, the robust critical value reduces to a chi-squared critical value asymptotically. 
If $H=V$, $\theta_\ast$ is in the interior of $\Theta$, and identification is strong, then $\widehat{CV}_L$ converges in probability to a chi-squared critical value. 
If, in addition, $\alpha_c=\alpha_\Psi=0$, then the RQLR test is asymptotically equivalent to the usual QLR test. 

\textbf{4.20.} While Theorem \ref{Theorem3} focuses on hypothesis testing, the RQLR test can also be used to construct confidence sets for $r(\theta)$ by inverting a family of tests. 
By incorporating sequences of null hypotheses as in \cite{AndrewsChengGuggenberger2020}, such confidence sets can be shown to have asymptotically uniformly valid coverage probability. 

\textbf{4.21.} The proof of Theorem \ref{Theorem3} is challenging because of the effect of the boundary under semi-strong identification. 
The proof strategy is to show that the estimated strong and weak quantiles adapt to the semi-strong quantiles; see Lemma G.3 in the supplemental materials. 
This strategy avoids explicitly accounting for and estimating the effect of the boundary under semi-strong identification, which would be very challenging because of the arbitrarily slow rate of the influence of the bounds. 
Overall, the RQLR test and the proof of Theorem \ref{Theorem3} provide a solution to the challenge of combining weak identification with informative bounds by handling the influence of the boundary in a way that is uniformly valid over the strength of identification. 
\qed
\medskip

To finish this section, we give instructions for simulating the quantiles of $QLR^L_\ast$ for $L\in\{S, W1, W2\}$. 
For simplicity, we state the instructions for $L=W2$ because the instructions for the other values of $L$ are similar. 
Suppose we have estimators $\hat\pi$, $\hat H$, $\hat V$, $\hat\pi_\ast\in\hat\Pi$, and $\hat\beta_\ast\in\mathcal{B}^r(\hat\pi)$. 
\begin{enumerate}
\item[Step 1:] Draw a large number, $B$, draws of $Y_b\sim N(0,\hat V)$ for $b\in\{1,...,B\}$. 
\item[Step 2:] For each $b$, calculate $\tilde q^W(\beta)$ and $\tilde q^{W,r}(\beta)$ as a function of $\beta\in\mathcal{B}(\hat\pi)$, using the formulas in (\ref{3.7}) and (\ref{4.2}), where $\pi_\ast$ is replaced by $\hat\pi$, $H$ is replaced by $\hat H$, and $c(\beta)$ is replaced by $\hat c(\beta; \hat\pi_\ast,\hat\beta_\ast)$, defined in (\ref{dform}). 
\item[Step 3:] For each $b$, calculate $QLR^{W2}_\ast$ by minimizing $\tilde q^W(\beta)$ over $\beta\in\mathcal{B}(\hat\pi)$ and by minimizing $\tilde q^{W,r}(\beta)$ over $\beta\in\mathcal{B}^r(\hat\pi)$. 
\item[Step 4:] Let the simulated $1-\alpha$ quantile of $QLR^{W2}_\ast$ be the $(1-\alpha)B$ order statistic of the values of $QLR^{W2}_\ast$. 
\end{enumerate}

\section{Simulations}

This section reports simulated rejection probabilities for hypothesis tests in Example 1. 
We find that the RQLR test is the only WIR test that can control size up to and including the boundary of the identified set and still have power against hypotheses that violate the bounds under non-identification. 

In Example 1, we test hypotheses on $\sigma^2$, the variance of the factor. 
Suppose the true values of the parameters are $\sigma^2=1$, $\lambda_2=1$, $\lambda_3=bn^{-1/2}$, and $\phi_j=1$ for $j\in\{1,2,3\}$. 
When $b=0$, $\sigma^2$ is not identified. 
It is only partially identified by bounds that characterize the identified set to be $[0.5, 2]$. 
In this case, by observational equivalence, the rejection probability should be less than or equal to $\alpha=0.05$ for every hypothesized value in the identified set. 
We simulate 1000 samples of size $n=500$ with iid normally distributed factors and errors and calculate rejection probabilities for tests of the hypothesis $H_0: \sigma^2=\sigma^2_0$, where $\sigma^2_0$ takes values in $[0.4,2.15]$. 

\begin{figure}[t]
\caption{Simulated Rejection Probabilities of Tests in Example 1: $b=0$} 
\pgfplotsset{width=0.5*\textwidth}
\begin{center}
\begin{tikzpicture}
\begin{axis}
[ymin=0,ymax=0.5,xmin=0.4,xmax=0.7,legend style={at={(axis cs:(.575,0.485)},anchor=north west}]
\draw[thick] (0.5,0.05) -- (0.7,0.05);
\addplot[very thick, dashed, purple] table [x=beta_0, y=AR-Plug, col sep=semicolon] {One_Factor_Sims_RP_b_0.txt};
\addplot[thick, densely dashdotted, green]  table [x=beta_0, y=AR-Plug-GEL, col sep=semicolon] {One_Factor_Sims_RP_b_0.txt};
\addplot[very thick, loosely dotted, magenta] table [x=beta_0, y=AM-Plug, col sep=semicolon] {One_Factor_Sims_RP_b_0.txt};
\addplot[thick, loosely dashed, red]  table [x=beta_0, y=BCS1, col sep=semicolon] {One_Factor_Sims_RP_b_0.txt};
\addplot[thick, dash dot dot, blue] table  [x=beta_0, y=AG-AR-Plug, col sep=semicolon] {One_Factor_Sims_RP_b_0.txt};
\addplot[very thick, dotted, teal] table  [x=beta_0, y=KMS_Rejection_Rate, col sep=semicolon] {One_Factor_Sims_KMS_output_b_0_converted.txt};
\addplot[very thick, densely dotted, violet]  table [x=beta_0, y=RQLR, col sep=semicolon] {One_Factor_Sims_RP_b_0.txt};
\addlegendentry{\footnotesize{AR}};
\addlegendentry{\footnotesize{GELR}};
\addlegendentry{\footnotesize{AM}};
\addlegendentry{\footnotesize{BCS}};
\addlegendentry{\footnotesize{SR-AR}};
\addlegendentry{\footnotesize{KMS}};
\addlegendentry{\footnotesize{RQLR}};
\end{axis}

\node[align=center, below] at (3,-.6) {(a) Lower Bound};
\end{tikzpicture}
\begin{tikzpicture}
\begin{axis}
[ymin=0,ymax=0.5,xmin=1.8,xmax=2.15,legend style={at={(axis cs:(1.805,0.485)},anchor=north west}]
\draw[thick] (1.8,0.05) -- (2,0.05);
\addplot[very thick, dashed, purple] table [x=beta_0, y=AR-Plug, col sep=semicolon] {One_Factor_Sims_RP_b_0.txt};
\addplot[thick, densely dashdotted, green]  table [x=beta_0, y=AR-Plug-GEL, col sep=semicolon] {One_Factor_Sims_RP_b_0.txt};
\addplot[very thick, loosely dotted, magenta] table [x=beta_0, y=AM-Plug, col sep=semicolon] {One_Factor_Sims_RP_b_0.txt};
\addplot[thick, loosely dashed, red]  table [x=beta_0, y=BCS1, col sep=semicolon] {One_Factor_Sims_RP_b_0.txt};
\addplot[thick, dash dot dot, blue] table  [x=beta_0, y=AG-AR-Plug, col sep=semicolon] {One_Factor_Sims_RP_b_0.txt};
\addplot[very thick, dotted, teal] table  [x=beta_0, y=KMS_Rejection_Rate, col sep=semicolon] {One_Factor_Sims_KMS_output_b_0_converted.txt};
\addplot[very thick, densely dotted, violet]  table [x=beta_0, y=RQLR, col sep=semicolon] {One_Factor_Sims_RP_b_0.txt};
\end{axis}

\node[align=center, below] at (3,-.6) {(b) Upper Bound};
\end{tikzpicture}
\end{center}
{\small
\begin{tablenotes}
\item {\em Note:} Finite sample ($n=500$) rejection probabilities of tests for $H_0: \sigma^2=\sigma^2_0$ in Example 1, calculated with 1000 simulations for each value. 
The true data generating process simulates jointly normally distributed errors and factors with $\sigma^2=1$, $\lambda_2=1$, $\lambda_3=0$, and $\phi_j=1$ for $j\in\{1,2,3\}$. 
\end{tablenotes}
}
\label{1F3M_1}
\end{figure}

Figure \ref{1F3M_1} focuses on the $b=0$ case and reports rejection probabilities for the Anderson-Rubin (AR) test from \cite{StockWright2000}, the GELR test from \cite{GuggenbergerSmith2005}, the likelihood ratio test (AM) from \cite{AndrewsMikusheva2016a}, the minimum resampling test (BCS) from \cite{BugniCanayShi2017}, the SR-AR test in \cite{AndrewsGuggenberger2019}, the calibrated projection test (KMS) from \cite{KaidoMolinariStoye2019}, and the RQLR test. 
We calculate subvector versions of the AR, GELR, AM, and SR-AR tests with strongly identified nuisance parameters plugged in.\footnote{We employ the reparameterization in Section 2, so all the nuisance parameters are strongly identified. This allows us to use plug-in versions of tests that are designed for full-vector inference. Otherwise, those tests would have to be projected.} 

\textbf{Remarks: 5.1.} Several WIR tests are not included in Figure \ref{1F3M_1}. 
The K and CLR tests defined in \cite{Kleibergen2005} reduce to the AR test when the model is just identified, as Example 1 is. 
Also, the versions of the AR, K, and CLR tests defined for minimum distance models in \cite{Magnusson2010} are equivalent to the GMM versions when the link function is additively separable between the reduced-form parameters and the structural parameters, as it is in Examples 1 and 2. 
We also omit projected versions of the WIR tests defined for full-vector inference, including the AR, K, and CLR tests because we expect them to be very conservative. 
Tests that explicitly allow for weakly identified nuisance parameters, including the tests in \cite{ChaudhuriZivot2011}, \cite{AndrewsMikusheva2016b}, \cite{Andrews2017}, and \cite{Andrews2018}, are compared in \cite{CoxWeakIdFactor}. 
\cite{CoxWeakIdFactor} found that those tests can be very conservative under weak identification relative to tests that reparameterize the model so the nuisance parameters are strongly identified. 
One could also consider WIR tests that ignore the bounds. 
Any such test must have trivial power against hypotheses that violate the bounds. 
Thus, the nontrivial power of the RQLR test quantifies the value of imposing the bounds. 

\textbf{5.2.} The solid black line denotes the $5\%$ threshold over the identified set. 
The RQLR test approximately controls size when $\sigma^2_0$ is in the identified set and has rejection probability above $5\%$ when $\sigma^2_0$ is outside the identified set. 
The AR, GELR, and AM tests have rejection probabilities above $5\%$ when $\sigma^2_0$ is in the identified set and therefore do not control size. 
Indeed, when $\sigma^2_0=2$, the maximal rejection probabilities over the identified set are $11.1\%$, $10.3\%$, and $28.5\%$, respectively. 
The SR-AR test rejects with probability approximately $5\%$ for any value of $\sigma^2_0$, so it does not have consistent power against hypotheses that violate the bounds. 

\textbf{5.3.} The problem with plugging in strongly identified nuisance parameters is that the estimators for the nuisance parameters are not asymptotically normal when the bounds are imposed. 
Instead, they have a parameter-on-the-boundary asymptotic distribution; see \cite{Andrews1999}. 
The argument used in \cite{StockWright2000}, \cite{Kleibergen2005}, \cite{AndrewsMikusheva2016a}, and others to show that strongly identified nuisance parameters can be plugged in relies essentially on asymptotic normality. 
Also, the formulas for the subvector test statistics in \cite{AndrewsGuggenberger2019} orthogonalize the moments away from the plug-in estimator of the strongly identified parameters. 
This makes the SR-AR test robust to the boundary but removes any power that can be gained from using the bounds. 

\textbf{5.4.} Figure \ref{1F3M_1} shows that the RQLR test is conservative, especially on the interior of the identified set. 
There are two reasons for this. 
(a) The use of $\alpha_\Psi$ and $\alpha_c$, combined with the Bonferroni argument, which is naturally conservative. 
(b) The critical value under weak identification takes the maximum over $\hat\pi_\ast$ and $\hat\beta_\ast$. 
This maximum is designed to control size for the least favorable asymptotic distribution, while the actual asymptotic distribution for this specification may be more favorable. 

\textbf{5.5.} Figure \ref{1F3M_1} includes BCS and KMS as points of comparison from the set-identification literature. 
(Example 1 can be cast as a moment equality model, which allows tests for moment inequalities to be implemented.) 
Even though these tests are not designed for weak identification, we can still evaluate their performance. 
\cite{Bei2024} proposed a computationally attractive version of the BCS test that is omitted because it should be similar to the BCS test. 
Also, methods designed for linear moments are omitted because Example 1 has a nonlinear moment. 
Figure \ref{1F3M_1} shows that BCS is more conservative than RQLR, which is surprising because BCS has not been shown to be valid under weak identification. 
It is unclear whether BCS is valid more generally for weakly identified models or if it only happens to be valid in this specification. 
Figure \ref{1F3M_1} shows that KMS is invalid with a maximal rejection probability over the identified set of 16.9\%. 
Why KMS is invalid is unclear. 
One possibility is that the rho-box construction includes a tuning parameter that has not been calibrated to have good finite-sample size in a weakly identified model. 

\textbf{5.6.} On average, the AR, GELR, AM, BCS, SR-AR, KMS, and RQLR tests take 0.02, 0.3, 248, 19, 0.03, 1667, and 43 seconds per simulation to compute, respectively.\footnote{The computation time for KMS is for the confidence interval, which should be divided by the number of grid points to be comparable. In this simulation, the number of grid points is 15.} 
The RQLR test is more computationally expensive than AR, GELR, BCS, and SR-AR, but is less computationally expensive than AM and KMS. 
See Section C.2 in the Supplemental Materials for details on the computation of these tests. 
\qed\medskip

\begin{figure}[t]
\caption{Simulated Power Functions of Tests in Example 1: $b>0$} 
\pgfplotsset{width=0.5*\textwidth}
\begin{center}
\begin{tikzpicture}
\begin{axis}
[ymin=0,ymax=0.8,xmin=0.4,xmax=1.6,legend style={at={(axis cs:(1.125,0.785)},anchor=north west}]
\draw (0.4,0.05) -- (1.6,0.05);
\addplot[thick, loosely dashed, red]  table [x=beta_0, y=BCS1, col sep=semicolon] {One_Factor_Sims_RP_b_4.txt};
\addplot[thick, dash dot dot, blue] table  [x=beta_0, y=AG-AR-Plug, col sep=semicolon] {One_Factor_Sims_RP_b_4.txt};
\addplot[very thick, densely dotted, violet]  table [x=beta_0, y=RQLR, col sep=semicolon] {One_Factor_Sims_RP_b_4.txt};
\addlegendentry{\footnotesize{BCS}};
\addlegendentry{\footnotesize{SR-AR}};
\addlegendentry{\footnotesize{RQLR}};
\end{axis}

\node[align=center, below] at (3,-.6) {(a) $b=4$};
\end{tikzpicture}
\begin{tikzpicture}
\begin{axis}
[ymin=0,ymax=0.8,xmin=0.6,xmax=1.4,legend style={at={(axis cs:(1.805,0.485)},anchor=north west}]
\draw (0.4,0.05) -- (1.6,0.05);
\addplot[thick, loosely dashed, red]  table [x=beta_0, y=BCS1, col sep=semicolon] {One_Factor_Sims_RP_b_16.txt};
\addplot[thick, dash dot dot, blue] table  [x=beta_0, y=AG-AR-Plug, col sep=semicolon] {One_Factor_Sims_RP_b_16.txt};
\addplot[very thick, densely dotted, violet]  table [x=beta_0, y=RQLR, col sep=semicolon] {One_Factor_Sims_RP_b_16.txt};
\end{axis}

\node[align=center, below] at (3,-.6) {(b) $b=16$};
\end{tikzpicture}
\end{center}
{\small
\begin{tablenotes}
\item {\em Note:} Finite sample ($n=500$) power functions of tests for $H_0: \sigma^2=\sigma^2_0$ in Example 1, calculated with 1000 simulations for each value. 
The true data generating process simulates jointly normally distributed errors and factors with $\sigma^2=1$, $\lambda_2=1$, $\lambda_3=bn^{-1/2}$, and $\phi_j=1$ for $j\in\{1,2,3\}$. 
\end{tablenotes}
}
\label{1F3M_2}
\end{figure}

We next evaluate the power of these WIR tests under strong and weak identification. 
We restrict attention to WIR tests that are valid in Figure \ref{1F3M_1}. 
Figure \ref{1F3M_2} reports power functions for the BCS, SR-AR, and RQLR tests when $\lambda_3=bn^{-1/2}$ with $b\in\{4,16\}$. 
The value of $b$ determines the strength of identification. 
Larger values of $b$ lead to stronger identification because the true value of the parameters is further away from the non-identified region of the parameter space. 

\textbf{Remark:} \textbf{5.7.} When $b=16$, the SR-AR test has highest power. 
It is asymptotically efficient under strong identification. 
The RQLR test is close, having slightly less power because it is also asymptotically efficient, except with size $\alpha-\alpha_\Psi=4.5\%$. 
The BCS test has the lowest power because it is asymptotically inefficient. 
When $b=4$, the power functions transition between the non-identified case and the strongly identified case. 
The power of the tests can generally be ranked in the same order as when $b=16$, except that the RQLR test has higher power for $\sigma^2_0<0.5$. 
This occurs because the RQLR test has more power against hypotheses that violate the bounds. 
Comparing the RQLR and SR-AR tests in Figures \ref{1F3M_1} and \ref{1F3M_2}, we can conclude that the RQLR test gives up some power against local alternatives under strong identification in exchange for power against global alternatives under weak identification. 
\qed\medskip

Section C.1 in the Supplemental Materials presents additional simulation results in a model with two factors. 

\section{Empirical Application}

The literature on childhood development commonly uses factor models to deal with measurement error in parental investments and skills. 
We demonstrate our WIR inference in the factor model used by \cite{Attanasio2020AER} to analyze the effects of a randomized intervention in Colombia. 

\cite{Attanasio2014} document a significant improvement in cognition and language development of young children following a randomized controlled trial of weekly home visits by trained local women designed to improve the quality of maternal-child interactions. 
\cite{Attanasio2020AER} analyze this intervention using a factor model for measures of parental time and material investments.\footnote{\cite{Attanasio2020AER} use the exogenous variation from the randomized controlled trial to identify parameters in the production function for human capital. We consider the more modest question of quantifying the effect of the intervention on parental investments.} 
The factors represent unobserved parental investments in children. 
The variance of the factors measures the heterogeneity of parental investments across the sample. 
A decrease in the variance of a factor, for example, indicates less dispersion in the heterogeneity of parental investments, possibly because the intervention was more effective in increasing parental investments for households that started with a lower level of parental investments. 
In particular, any change in the variance of a factor is inconsistent with a homogeneous treatment effect. 

The factor model used by \cite{Attanasio2020AER} specifies one factor for the control group and one factor for the treatment group.\footnote{\cite{Attanasio2020AER} ran a variety of exploratory factor analyses on the measurement system for parental investments that includes both time and material investments and found anywhere between 1 and 4 factors; see Table C.1 in \cite{Attanasio2020AER}.} 
\cite{CoxWeakIdFactor} calculates a specification test for the one-factor models for material investments and notes that they are likely misspecified with p-values below $10^{-8}$. 
To account for this, \cite{CoxWeakIdFactor} estimates factor models with two factors, allowing the second factor to be weakly identified. 

To be specific, suppose the dataset contains $J$ measures of material investments in children. (These are just variables indicating the number and type of toys the children have; see the supplemental materials for more details on the dataset and factor model specification.) 
Control and treatment will be abbreviated by c and t in what follows. 
Individual $i$ with treatment status $s\in\{c,t\}$ has observed material investment measure $j\in\{1,...,J\}$ given by 
\begin{equation}
X^s_{ij}=\mu^s_j+{\lambda^s_j}' f^s_i+\epsilon^s_{ij}, 
\end{equation}
where $\mu^s_j$ is the mean of measure $j$, $\lambda^s_j$ are the factor loadings for measure $j$, $f_i^s$ are the parental investment factors, and $\epsilon_{ij}^s$ are error terms assumed uncorrelated with each other and with the factors. 
Let $\sigma^2_{s,1}$ and $\sigma^2_{s,2}$ denote the variances of the two factors for $s\in\{c,t\}$. 

A factor loading is positive if purchasing a particular type of toy (say, coloring books) constitutes an investment in the skills of the child. 
This justifies the maintained assumption that four of the factor loadings are nonzero. 
Even if all the factor loadings are positive, there can still be an identification problem if all the toys constitute an investment in the same linear combination of skills. 

\cite{CoxWeakIdFactor} compares WIR hypothesis tests in low-dimensional factor models and recommends a test that uses a reparameterization so the nuisance parameters are strongly identified, such as the CLR test. 
Table \ref{EmpiricalResults} reports point estimates and confidence intervals (CIs) for the variances of the factors, including a standard CI calculated with plug-in standard errors, the CLR CI that inverts the CLR test, and the RQLR CI that inverts the RQLR test. 
We make the following remarks on Table \ref{EmpiricalResults}. 

\begin{table}[t]
\begin{center}
\scalebox{\shrinkageparameter}{
\begin{threeparttable}
\caption{Parental Investment Factors}\label{EmpiricalResults}
\begin{tabular}{lccccc}
\hline\hline\vspace{-0.3cm}\\
&\multicolumn{2}{c}{Control}&&\multicolumn{2}{c}{Treatment}\\
\cline{2-3}\cline{5-6}\vspace{-0.3cm}\\
&{$\sigma^2_{c,1}$}&{$\sigma^2_{c,2}$}&&{$\sigma^2_{t,1}$}&{$\sigma^2_{t,2}$}\\
\hline\vspace{-0.3cm}\\
Point Estimate&0.93&0.32&&1.00&0.08\\
Standard CI&[0.80,1.06]&[0.20,0.44]&&[0.83,1.18]&[0.04,0.12]\\
Standard CI Length&0.26&0.24&&0.35&0.08\\
CLR CI&[0.90,1.34]&[0.17,0.34]&&[0.98,10]&[0.03,0.08]\\
CLR CI Length&0.44&0.17&&9.02&0.05\\
RQLR CI&[0.84,0.95]&[0.25,0.40]&&[0.88,1.10]&[0.08,0.14]\\
RQLR CI Length&0.11&0.15&&0.22&0.06\\
\hline
\end{tabular}
{\small
\begin{tablenotes}
\item {\em Note:} The point estimate minimizes the minimum distance objective function. 
The standard CI is calculated using a t-statistic. 
The CLR CI is calculated by inverting the CLR-Plug test without imposing the bounds. 
The RQLR CI is calculated by inverting the RQLR test. 
\end{tablenotes}
}
\end{threeparttable}
}
\end{center}
\end{table}

\textbf{Remarks. 6.1.} The Standard CI is only valid under strong or semi-strong identification. 
The CLR CI is calculated by inverting a plug-in version of the CLR test from \cite{Kleibergen2005} using a reparameterization to make the nuisance parameters strongly identified. 
The CLR statistic is calculated using a version of the AR and K statistics that does not impose the bounds. 
This is because the simulations in Section 5 indicate that the version of the CLR test that does impose the bounds is invalid near the ends of the identified set. 
The RQLR CI is calculated by inverting the RQLR test. 

\textbf{6.2.} From Table \ref{EmpiricalResults}, we see that the variance of the first factor is roughly the same between the treatment and control groups. 
This indicates that the treatment had little effect on the heterogeneity of the first parental investment, possibly because the treatment effect is homogenous. 
The variance of the second factor, however, is lower for the treatment group, possibly because the treatment is more effective in increasing the second parental investment for households that started with a lower level. 
This emphasizes the importance of allowing for two types of parental investments. 

\textbf{6.3.} The CLR CI for $\sigma^2_{t,1}$ is very wide, with right endpoint at 10. 
This is because 10 was used as an arbitrary upper bound on the values of these parameters for practical/numerical reasons. 
If this number were increased, the right endpoint would likely be larger and possibly unbounded. 
This is a common problem with WIR CIs.\footnote{Without any bounds, \cite{Dufour1997} shows that any CI that is valid under non-identification must have infinite length with probability at least $1-\alpha$ asymptotically. This result no longer applies when bounds are used.} 
In factor models, this is especially problematic because the variance of a factor cannot be larger than the variance of its normalizing variable. 
For $\sigma^2_{t,1}$, the point estimate of the upper bound is $1.01$. 
Imposing the bounds and using the RQLR test is a remedy to this problem. 
The RQLR CI is reasonable in length and does not extend much beyond the point estimate of the upper bound. 
\qed

\section{Conclusion}

This paper combines weak identification with bounds in a class of minimum distance models. 
Limit theory for the minimum distance estimator and the QLR statistic is derived for strong, semi-strong, and weak sequences of parameters. 
An identification-robust test (the RQLR test) that has power against hypotheses that violate the bounds is recommended for testing hypotheses on the structural parameters. 
The RQLR test is implemented in two example factor models with weak factors and an empirical application, demonstrating the value of bounds as a source of information when parameters are weakly identified. 

\section*{Acknowledgements}

The author is grateful for helpful comments received from the co-editor, the associate editor, two anonymous referees, and the participants of seminars/conferences at Brown, Columbia, Duke, LSE, NUS, Penn, Princeton, UCL, University of Amsterdam, University of Rochester, and Yale. 
The author is especially grateful for comments from Xu Cheng, Adam McCloskey, and Sukjin Han, as well as advice and guidance from Donald Andrews, Xiaohong Chen, Yuichi Kitamura, and Serena Ng.

\begin{singlespace}
\bibliography{references}
\end{singlespace}

\end{document}